\documentclass[10pt, twocolumn, notitlepage, prb, superscriptaddress, longbibliography]{revtex4-2}

\usepackage{here}
\usepackage{amsmath}         
\usepackage{graphicx}
\usepackage{braket}         
\usepackage{lettrine}
\usepackage{xcolor}
\usepackage{hhline}
\usepackage{tabularx}
\usepackage{calc}
\usepackage{soul}
\usepackage{array}
\usepackage[utf8]{inputenc}
\DeclareUnicodeCharacter{030C}{\v{}}
\DeclareUnicodeCharacter{0301}{\'}
\DeclareUnicodeCharacter{0303}{\~{}}
\usepackage[colorlinks=true,linkcolor=blue, citecolor=blue, urlcolor=blue]{hyperref}%

\DeclareTextFontCommand{\DG}{\color{red}\bfseries} 

\DeclareTextFontCommand{\MZ}{\color{green}\bfseries} 

\begin{document}

\title{Orbital torques and orbital pumping in two-dimensional rare-earth  dichalcogenides} 

\author{Mahmoud Zeer}
\email{m.zeer@fz-juelich.de}
\affiliation{Peter Gr\"unberg Institute, Forschungszentrum J\"ulich, 52425 J\"ulich, Germany}
\affiliation{Department of Physics, RWTH Aachen University, 52056 Aachen, Germany}
\affiliation{Institute of Physics, Johannes Gutenberg-University Mainz, 55099 Mainz, Germany}

\author{Dongwook Go}

\affiliation{Peter Gr\"unberg Institute, Forschungszentrum J\"ulich, 52425 J\"ulich, Germany}
\affiliation{Institute of Physics, Johannes Gutenberg-University Mainz, 55099 Mainz, Germany}

\author{Mathias Kl\"aui}
\affiliation{Institute of Physics, Johannes Gutenberg-University Mainz, 55099 Mainz, Germany}

\affiliation{Centre for Quantum Spintronics, Department of Physics, Norwegian University of Science and Technology, 7491 Trondheim, Norway}

\author{Wulf Wulfhekel}
\affiliation{Physikalisches Institut, Karlsruhe Institute of Technology, 76131 Karlsruhe, Germany}

\author{Stefan Bl\"ugel}
\affiliation{Peter Gr\"unberg Institute, Forschungszentrum J\"ulich, 52425 J\"ulich, Germany}

\author{Yuriy Mokrousov}
\email{y.mokrousov@fz-juelich.de}
\affiliation{Peter Gr\"unberg Institute, Forschungszentrum J\"ulich, 52425 J\"ulich, Germany}
\affiliation{Institute of Physics, Johannes Gutenberg-University Mainz, 55099 Mainz, Germany}

\begin{abstract}
The design of spin-orbit torque properties in two-dimensional (2D) materials presents one of the challenges of modern spintronics. In this context,  2D layers involving rare-earth ions $-$ which give rise to robust magnetism, exhibit pronounced orbital polarization of the states, and carry strong spin-orbit interaction $-$ hold particular promise. Here, we investigate ferromagnetic Janus H-phase monolayers of 4$f$-Eu rare-earth dichalcogenides EuSP, EuSSe, and EuSCl using first-principles calculations.
We demonstrate that all compounds exhibit significant spin-orbit torques which originate predominantly in the colossal current-induced orbital response on the Eu $f$-electrons. Moreover, we demonstrate that the corresponding orbital torques can be used to drive strong in-plane currents of orbital angular momentum with non-trivial direction of orbital polarization. Our findings promote $f$-orbital-based 2D materials as a promising platform for in-plane orbital pumping and spin-orbit torque applications, and motivate further research on educated design of orbital properties for orbitronics with 2D materials.
\end{abstract}

\date{\today}                 
\maketitle	      

\section*{Introduction}
Orbitronics, holding significant potential to shape information technology, has emerged as a field capable of developing eco-friendly electronic devices~\cite{shao2021roadmap}. By harnessing the unique properties of electron's orbital angular momentum (OAM), which can be generated in a highly efficient manner by the direct transfer of angular momentum from the lattice, 
orbitronics opens up new possibilities for energy-efficient technologies that can reduce environmental impact and optimize device performance~\cite{jo2024spintronics}. The study of OAM offers the opportunity to explore novel data storage and processing mechanisms, which could revolutionize modern computing \cite{wang2024orbitronics,xu2024orbitronics}. The OAM is the pivotal degree of electronic freedom, which can result in various phenomena such as the orbital Hall effect (OHE) \cite{Tanaka2008, Go2018}, orbital magnetoelectric effects, and orbital torques \cite{go2020orbital,hayashi2023observation,hayashi2023observation,go2023long}. In this context, orbital torque and orbital pumping emerge as critical mechanisms in orbitronics. Orbital torque, akin to spin torque in spintronics ~\cite{lee2021orbital,hayashi2023observation,gambardella2011current}, allows for efficient manipulation of magnetization via the transfer of angular momentum. Similarly, orbital pumping facilitates the generation of orbital currents through dynamic coupling between orbitals and magnetization. Together, these mechanisms bridge theoretical insights with practical applications, advancing the potential of orbitronics in modern technology.

In broad terms, the current-induced magnetization dynamics involves the transfer of angular momentum, extending beyond spin alone. The key role of the OAM as a source of torque on the magnetization has been firmly established in the past years, both theoretically and experimentally~\cite{lee2021orbital,dutta2022observation,lee2021efficient,ding2020harnessing, hayashi2023observation}. The fundamental mechanism of exchange among the non-equilibrium OAM and spin via the process of spin-orbit coupling (SOC) mediated spin-orbital, or simply, {\it orbital torque}, has been recently solidified from quantum mechanical formulations and first principles calculations~\cite{Go2020theory}. As a result, this provided a key to understanding of various experiments where the excess of OAM is believed to be generated via the mechanisms of OHE or orbital Rashba-Edelstein effect~\cite{ding2022observation,krishnia2024quantifying}. In such cases, strong spin-orbit torques on the magnetization are generated, and they often exhibit an unusual behavior distinctly different from that expected from purely spin-mediated torques~\cite{manchon2019current,gupta2024generation}. 

Reciprocal to spin-orbit torque process, the magnetization dynamics generates a flow of charge, with the correlation between the direct and inverse process governed by the very same spin-orbit torkance tensor~\cite{freimuth2015direct}. Given the observed role of OAM in mediating the spin-orbit torque, it was postulated that, in addition to the effect of spin pumping, the magnetization dynamics is able to create an excess of OAM~\cite{go2023orbital, han2024orbital}, with the corresponding effect of {\it orbital pumping} observed very recently~\cite{hayashi2024observation}. Despite clear successes in promoting and exploiting the OAM for magnetization dynamics, still very little is  understood concerning the basic ingredients of the orbital torque and orbital pumping. In this context, the design of material families where various contributions to the spin-orbit torque and pumping can be suppressed and promoted in an educated way, is pertinent. In a recent work, Ding {\it et al.}~\cite{ding2024orbital} demonstrated that tuning the concentration of heavy Gd in a GdCo alloy can be used to effectively enhance the strength of orbital-to-spin conversion which has direct impact on the magnitude and sign of the torque. Strong spin-to-orbital conversion leading to THz emission has been also recently demonstrated in heterostructures containing Nd, Gd and Ho~\cite{liu2024qualitative}. This suggests that employing heavy  rare-earth elements for orbital torque and orbital current design can aid in gaining deeper insights into the interplay of spin and orbital degrees of freedom out of equilibrium. 

\begin{figure*}[t!]
    \includegraphics[angle=0, width=0.9\textwidth]{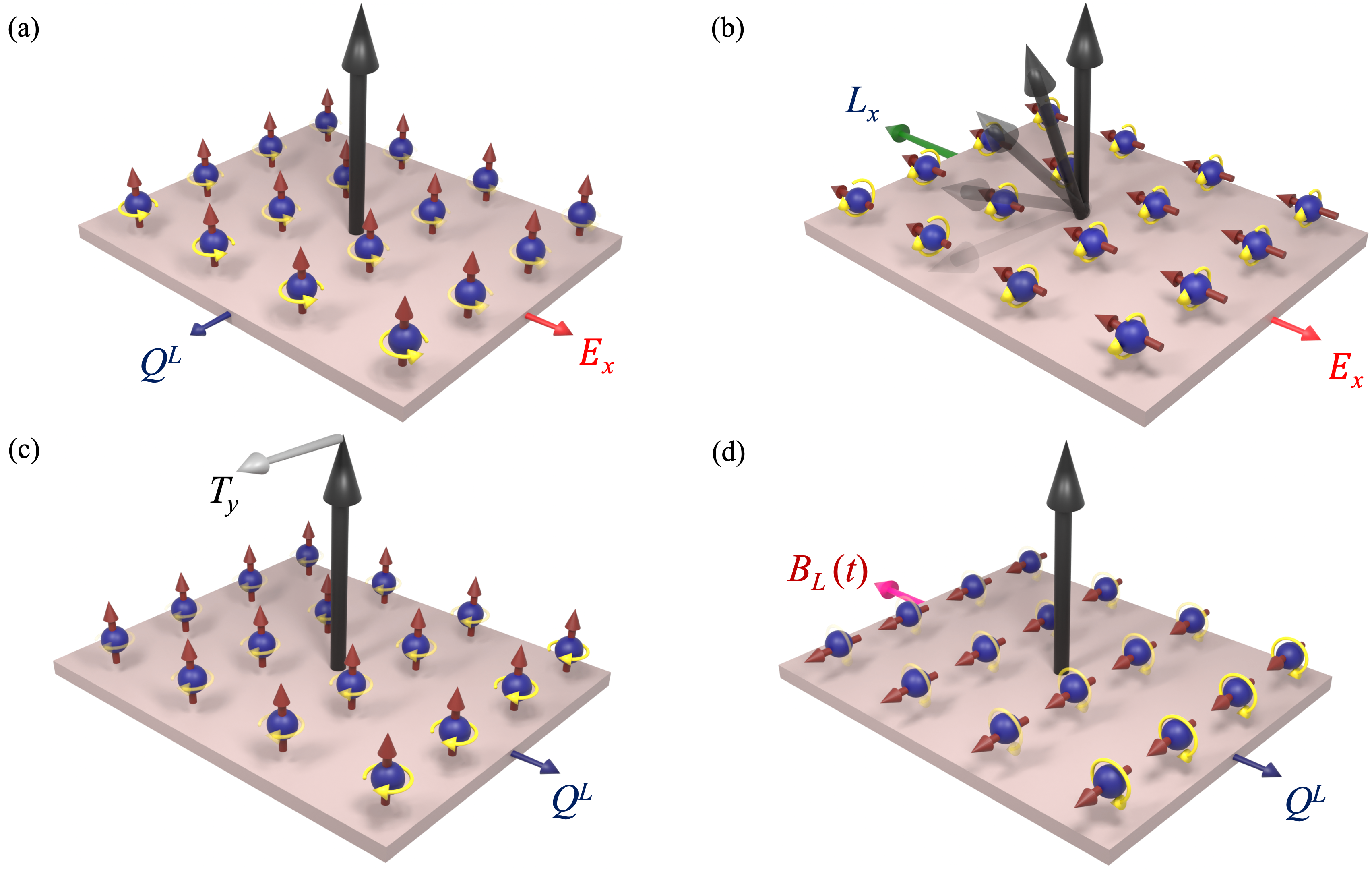}
    \caption{{   \bf Orbital torque and orbital pumping in 2D Janus dichalcogenides.} (a) In 2D magnetic dichalcogenides (magnetization out of the plane along $z$, indicated with a black arrow), application of an electric field $E_x$ along $x$ (red arrow) leads to generation of a current of orbital angular momentum $Q^L$ along $y$, polarized along $z$ (small red arrows and yellow circular arrows) via the mechanism of the {\it orbital Hall effect} (OHE). (b) At the same time, due to broken $z$-reflection symmetry in Janus geometry, the electric field will give rise to a current-induced orbital polarization $L_x$ (green arrow, as well as small red arrows and yellow circular arrows), which mediates an {\it orbital torque} on the magnetization along $y$ (depicted with fading arrows). (c) In a reciprocal process, perturbing the magnetization with a torque $T_y$ along $y$ (grey arrow) will result in a generation of an orbital current along $x$ polarized along $y$ and $z$ (shown with small yellow and red arrows), which constitutes the effect of {\it in-plane orbital pumping}. (d) For rare-earth dichalcogenides, the effect of orbital pumping is equivalent to the effect of {\it orbital-to-orbital-current conversion}, where the orbital current arises in response to an increasing  linearly in time orbital field $B_L(t)$ applied along $x$ (pink arrow). While the OHE can be used to generate orbital currents in 2D materials, the effect of orbital pumping mediated by orbital torque can be used to achieve  orbital current generation by magnetization dynamics in 2D geometry.} 
    \label{fig:pumping}
\end{figure*}
Among novel material classes, two-dimensional (2D) materials present an exciting platform for various novel phenomena, such as the OHE, valley-dependent orbital transport, and spin-orbit torque effects~\cite{gibertini2019magnetic,yi2019recent,ahn20202d,husain2020emergence}. The confined geometry and intrinsic anisotropy of these compounds especially promotes the orbital degree of freedom~\cite{grytsiuk2020topological,niu2019mixed}, which makes 2D materials a fruitful ground for testing and realizing various flavors of orbital effects.
For example, it was found that in transition-metal dichalcogenides such as MoS$_2$ and WS$_2$ the OHE, see Fig.~\ref{fig:pumping}(a), acquires exotic properties, and can even exhibit quantization~\cite{cysne2021disentangling,canonico2020orbital}, leading to a formulation of a concept of orbital Chern insulator~\cite{fan2024orbital,cysne2021disentangling}. It was shown that the valley degree of freedom which naturally emerges in transition-metal dichalcogenides~\cite{xue2020imaging,lee2016electrical,mak2014valley} is closely correlated with orbital transport out of equilibrium~\cite{cysne2021disentangling,bhowal2021orbital}. Additionally, it was demonstrated that utilizing rare-earth elements such as Gd and Eu to form 2D layers of rare-earth metal dichalcogenides, can give rise to strong magnetism, promote large orbital Hall response of $p$- and $f$-electrons, and result in robust quantum spin, anomalous and orbital Hall phases~\cite{zeer2022spin,zeer2024promoting}.

The design of spin-orbit torque in two dimensions is one of the actively explored areas of 2D magnetism~\cite{hidding2020spin}. Following the continuity equations for the redistribution of angular momentum, it has been suggested that in purely 2D systems the spin-orbit torque should be entirely attributed to the orbital contribution~\cite{Go2020theory,manchon2009theory}. It was also found that in 2D layers of finite thickness the spin and orbital nature of the torque can be controlled by the magnetization direction~\cite{saunderson2022hidden}. In particular, the so-called Janus 2D layers have attracted significant attention recently~\cite{guo2018,zhang2020recent} due to the combined properties of low-dimensionality, anisotropy, valley-spin coupling, and the Rashba coupling. The hexagonal phase of Janus layers has been predicted and experimentally realized in nonmagnetic TMDs MoSSe and WSSe~\cite{zhang2017,lu2017,zhang2020}. 
Low symmetry of these materials promotes phenomena such as the spin Hall effect~\cite{Yu2021} and the quantized transport due to nontrivial $\mathbf{Z}_2$ band topology~\cite{joseph2021,guo2021}.  The Janus layers were found to possess a high level of electrical conductivity, mechanical strength~\cite{van2020,shi2018}, and thermal stability~\cite{luo2021}, making them a promising material class for applications in electronics, energy storage, and spintronic devices \cite{tang2022,zhang2022,rezavand2022electronic}. The Janus structure has also attracted significant research in the context of spin-orbit torque switching~\cite{vojacek2024field}. 

In this work, taking as an example low-symmetry rare-earth dichacolgenides of Janus type, which combine strong magnetism with large SOC and orbital polarization, we explore the role of orbital magnetism in mediating spin-orbit torque and orbital pumping in 2D magnets based on $f$-electrons. First of all, we demonstrate that the current-induced torque acting on the $f$-states can reach an order of magnitude observed in most optimal representatives of the 2D magnets ~\cite{liu2020two,tang2021spin}. We find that the $f$-torque is purely {\it orbital} in nature, that is, it is  directly proportional to the current-induced local orbital moment of rare-earth atoms, Fig.~\ref{fig:pumping}(b). Second, we predict that igniting magnetization dynamics by external means will result in the effect of two-dimensional 
{\it orbital pumping}, with the pumped in-plane currents of angular momentum dominated by the orbital contribution, Fig.~\ref{fig:pumping}(c). Moreover, our prediction is that, due to the low symmetry of the structure, the response of pumped current is non-trivial in terms of direction and polarization, which may find direct applications in spintronic architectures based on an in-plane spin and OAM injection. Finally, we uncover the origin of the orbital pumping in the effect of {\it orbital-to-orbital-current conversion} mechanism, Fig.~\ref{fig:pumping}(d), where a perturbation of the Hamiltonian by the time-dependent orbital field  can fully replace the dynamical term in predicting the response, thus postulating an equivalence between the torque and OAM operators within the subspace of $f$-electrons. At the end of the manuscript, we discuss the implications of our findings in the discipline of 2D magnetism and spintronics research and experimental methods to confirm our predictions.  

\section*{Results}


\begin{figure*}[t!]
    \includegraphics[angle=0, width=0.98\textwidth]{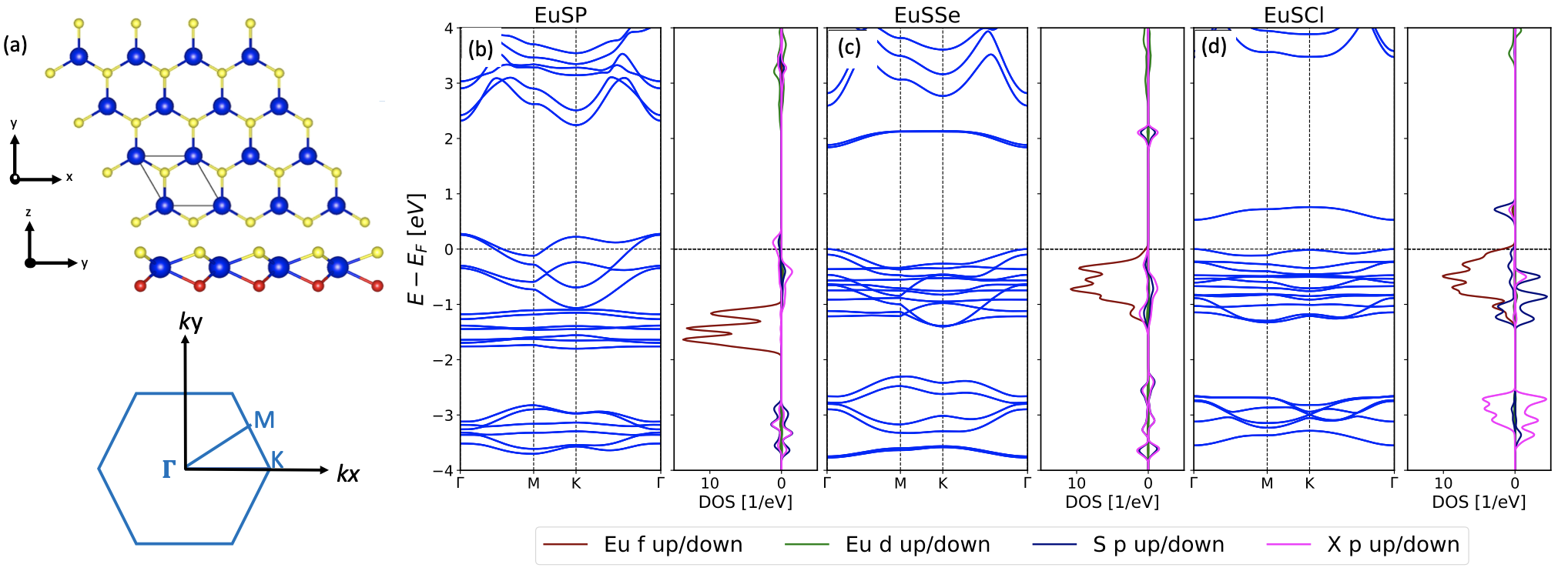}
    \caption{{\bf Electronic structure of EuSX}. (a) Top and side views of the Janus H-phase  EuSX monolayers (X = P, Se, and Cl) are shown. The dark blue balls represent Eu atoms, and the yellow/red balls represent S and X atoms, respectively, with the axes shown with black arrows. The first Brillouin zone is shown at the bottom. Right:   
    (b-d) The band structures and the corresponding spin-resolved density of states (DOS)  are shown. The left and right parts of the DOS correspond to the majority and minority spin, respectively. The DOS of Eu-$f$, Eu-$d$, S-$p$ and X-$p$ states is shown with dark red, green, blue and pink, respectively. 
}
    \label{fig:EuSX_bandDOS}
\end{figure*}

 \textbf{Electronic and magnetic properties}. We summarize in Table \ref{table:Lattice} of the Methods section the optimized 
structure lattice parameters, 
magnetic moments of Eu, S, and X atoms, and the band gaps calculated for considered systems with the magnetization along $z$, which we find to be the easy axis for all systems. Here, we consider only the case of out-of-plane magnetized EuSX, leaving the aspect of anisotropy of various effects with respect to the magnetization direction for future work.  We find that for all cases the Eu spin moment is close to 7\,$\mu_B$, reflecting the half-filled $f$-shell configuration ($4f^7 6s^2$), while X spin moment is opposite in sign, ranging in magnitude between 0.02\,$\mu_B$ for Se and 0.64\,$\mu_B$ for Cl. The orbital moments are always negative with respect to Eu spin moment, ranging between $-0.01\,\mu_B$ and $-0.19$\,$\mu_B$ for all atoms. 

We first analyze the evolution of the electronic structure and magnetic properties when increasing the number of electrons in the system by going from  EuSP, to EuSSe, and EuSCl. Note that the valence shell configurations for P, S, and Cl are $3s^2 3p^3$, $3s^2 3p^4$, and $3s^2 3p^5$, respectively. The bandstructures and spin-resolved density of states (DOS) for  EuSX monolayers computed with SOC are shown in Fig.~\ref{fig:EuSX_bandDOS}. 
Starting with the case of EuSP, Fig.~\ref{fig:EuSX_bandDOS}(b) we find that the system is a metal with states at the Fermi energy dominated by $p$-orbitals of P, and Eu-$f$ states located 1-2\,eV below. 

Remarkably, the $f$-states of Eu and the $p$-states of S or P atoms are always separated in energy, so that their hybridization is minimal.
In contrast, the $p$-states of S and P atoms exhibit a strong hybridization, particularly between $-$3\,eV and $-$4 eV.
Turning to EuSSe, Fig.~\ref{fig:EuSX_bandDOS}(c), we observe that it is a semiconductor with a large band gap  of 1.84\,eV. The valence band maximum and the conduction band minimum are both located at the $\Gamma$ point, with the former dominated by the 4$f$ majority states of the Eu atom, and the latter originating in the $p$ states of S/P atoms, with both orbital types hybridizing strongly near the Fermi energy. 
 
The results for the case of EuSCl are shown in 
Fig.~\ref{fig:EuSX_bandDOS}(d). This system has a direct semiconducting band gap of 0.534\,eV at $\Gamma$, which is generated by the flat band of majority 4$f$-state of the Eu atom and the minority 3$p$ of the Cl atom. It also exhibits 
the largest in magnitude spin moment of the S atom among all layers of 0.64\,$\mu_\mathrm{B}$, originating in the partial occupation of spin-split flat S $p$-bands at the Fermi energy. In contrast to other systems, there is a clear imbalance in the hybridization of the $p$- and $f$-states: Eu states bond much stronger with the $p$-states of S, while the group of states at around $-$3\,eV is predominantly of Cl character. The latter group of $p$-states also looks qualitatively different from the other two compounds. 

\vspace{0.2cm}
\textbf{Orbital torque in  EuSX.} The geometry of EuSX is of Janus type, where the structural inversion symmetry and the mirror reflection symmetry with respect to $z$ ($\mathcal{M}_z$) are naturally broken, which allows for the effect of spin-orbit torque for the out of plane magnetization. 

To investigate the torque that is exerted on the magnetization upon application of an electric field, we utilize the spin continuity equation for the time evolution of spin density~\cite{Go2020theory}:
\begin{equation}
\frac{\partial\mathbf{S}}{\partial t}={\Phi}^{\mathbf{S}}+{T}_{\rm SO}^\mathbf{S}+{T}_{\rm XC}^\mathbf{S},
\end{equation}
where $\boldsymbol{\Phi}^{\mathbf{S}}$ stands for the spin flux contribution, $\boldsymbol{T}_{\rm SO}$ is what we refer to as the spin-orbital, or, simply orbital torque, and $\boldsymbol{T}_{\rm XC}$ is the torque on the magnetization due to transfer of angular momentum from the spin of excited electronic states.

For the total Hamiltonian $H$, the spin flux coming into an atom defined by the projection $P$ is given by
\begin{equation}
{\Phi}^\mathbf{S} = 
\frac{1}{2i\hbar}
\left\{ 
[P, H], \ \mathbf{S}
\right\},
\end{equation}
where $[A,B]=AB-BA$ and $\{A,B\}=AB+BA$ are the commutator and anti-commutator for the pair of operators $A$ and $B$. The spin-orbital torque and exchange torque are given as
\begin{eqnarray}
{T}_{\rm SO}^\mathbf{S} = \frac{1}{2i\hbar}
\left\{ 
[\mathbf{S},H_\mathrm{SO}],\ P
\right\},
\\
{T}_{\rm XC}^\mathbf{S} = \frac{1}{2i\hbar}
\left\{ 
[\mathbf{S},H_\mathrm{XC}],\ P
\right\},
\end{eqnarray}
respectively, where $H_{\rm SO}$ and $H_{\rm XC}$ stand for the spin-orbit and spin-dependent exchange-correlation part of the single-particle Kohn-Sham Hamiltonian, respectively. That is, the total Hamiltonian is given by
\begin{equation}
H = H_0 + H_{\rm SO} + H_\mathrm{XC},
\end{equation}
where $H_0$ incorporates the spin-independent part of the Hamiltonian, including the kinetic energy and the crystal-field potential generated by the lattice and the mean-field of electron cloud. Note that the exchange Hamiltonian is given as $H_\mathrm{XC}=(2\mu_\mathrm{B}/\hbar) \mathbf{S}\cdot\boldsymbol{V}_{\rm XC}$, where $\boldsymbol{V}_{\rm XC}=\boldsymbol{\hat{M}}\cdot V_{\rm XC}(\mathbf{r})$ is the so-called exchange field which determines the magnitude of the magnetic moments and spatial properties of the spin density, and ${\boldsymbol{\hat{M}}}$ is the unit vector along the magnetization direction. 
 
In the steady state, the torque on the magnetization is given by
\begin{equation}
\left\langle -{T}_{\rm XC}^\mathbf{S} \right\rangle \approx \left\langle {\Phi}^{\mathbf{S}} \right\rangle + \left\langle {T}_{\rm SO}^\mathbf{S} \right\rangle,
\end{equation}
where $\langle \cdots \rangle$ stands for the statistical average in the steady state.  In a purely 2D system, periodic in the $xy$ plane,  the flux is absent by construction, and the exchange torque is purely given by the spin-orbital torque, the angular momentum transfer from the OAM to the spin, $\left\langle -{T}_{\rm XC}^\mathbf{S} \right\rangle \approx  \left\langle {T}_{\rm SO}^\mathbf{S} \right\rangle$. Under an external electric field $E_x$ applied along $x$, the response of the torque in the steady state is given by the Kubo formula

\begin{eqnarray}
\label{eq:Kubo_torque}
\left\langle T_\mathrm{XC}^\mathbf{S} \right\rangle 
&=&
\frac{e\hbar E_x}{N_{{\boldsymbol{k}}}} \sum_{nm} \sum_{\boldsymbol{k}} (f_{n\boldsymbol{k}} - f_{m\boldsymbol{k}})
\nonumber
\\
& & \ \
\times 
\frac{
\mathrm{Im}
\left[ 
\bra{\psi_{n\boldsymbol{k}}} T_\mathrm{XC}^\mathbf{S} \ket{\psi_{m\boldsymbol{k}}}
\bra{\psi_{m\boldsymbol{k}}} v_x \ket{\psi_{n\boldsymbol{k}}}
\right]
}{
(E_{n\boldsymbol{k}} - E_{m\boldsymbol{k}}+i\eta )^2}
\\
&\equiv & t_{yx} E_x,
\label{eq:torkance_definition}
\end{eqnarray}
where $e> 0$ is the unit charge, $N_{\boldsymbol{k}}$ is the number of $k$-points used for the summation such that the response is given in the unit per unit cell, $v_{x}$ is the $x$-component of the velocity operator, $\psi_{n\boldsymbol{k}}$ is an eigenstate of the total Hamiltonian with band index $n$,  $E_{n\boldsymbol{k}}$ is its energy eigenvalue, and $t_{yx}$ is the $yx$-component of the torkance tensor. Here, $f_{n\boldsymbol{k}}$ is the corresponding Fermi-Dirac distribution function, where we set the temperature $T=300\ \mathrm{K}$. The broadening parameter $\eta$ is effectively describes the degree of disorder, which also help numerical convergence. We set $\eta=25\ \mathrm{meV}$ in all the calculations shown below. The calculation for the spin-orbital torque is done by replacing $T_\mathrm{XC}^\mathbf{S}$ by $T_\mathrm{SO}^\mathbf{S}$ in Eq.~\eqref{eq:Kubo_torque}.  In the following, for brevity, we skip the explicit reference to the distibution function in the summation, referring thus to the zero temperature limit expressions for computed quantities.

\begin{figure*}[]
   \includegraphics[angle=270, width=0.99\textwidth]{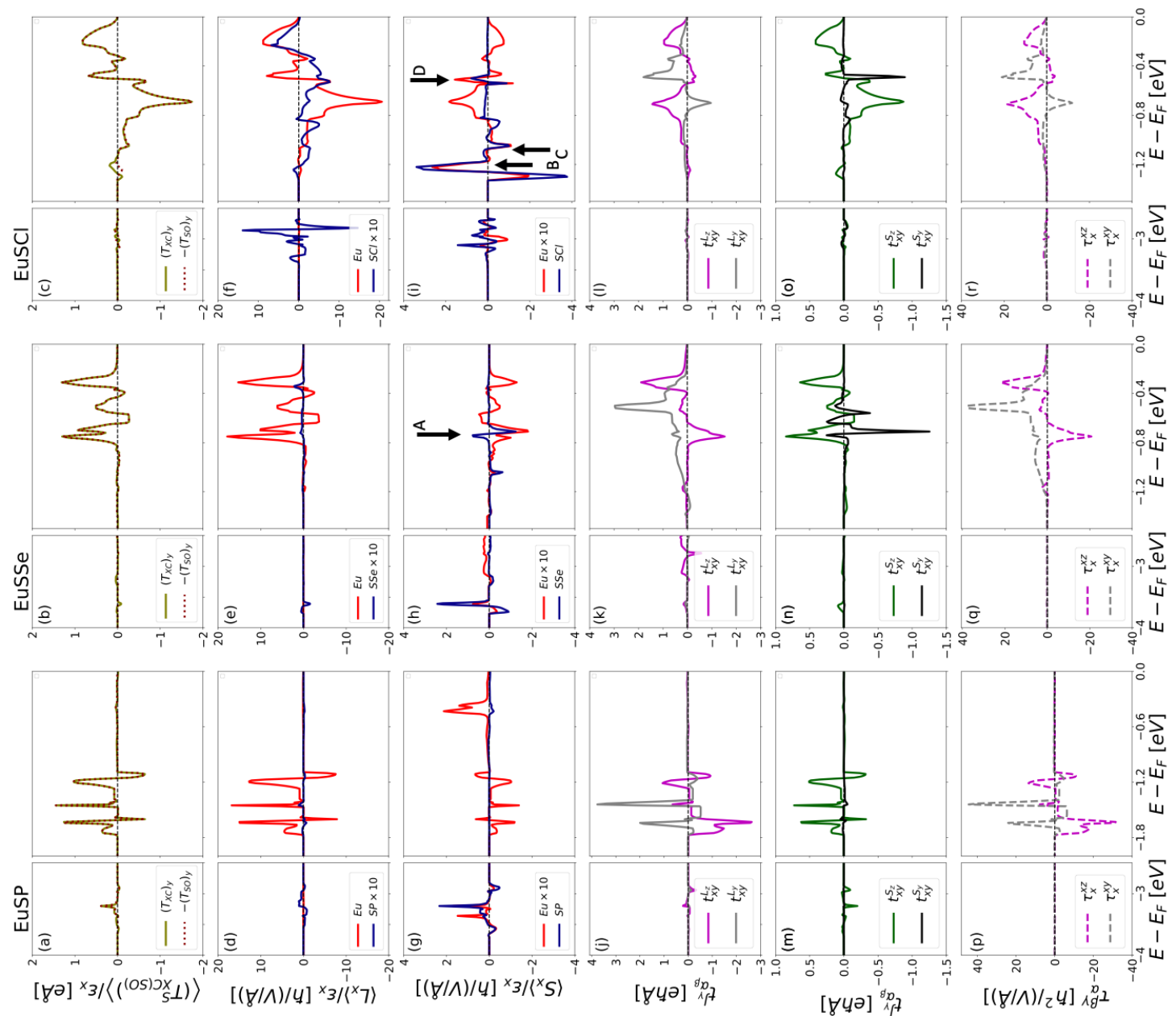}
   \caption{  {\bf Orbital torque and orbital pumping in EuSX.} (a-c) Band-filling dependence of the $y$-component of spin-orbital ($\langle T_{\rm SO}^{\mathbf{S}}\rangle$) and exchange ($\langle T_{\rm XC}^{\mathbf{S}}\rangle$) torque normalized to the strength of an electric field $E_x$ applied along $x$. The band-filling of the $x$-component of the normalized current-induced OAM ($\langle L_x\rangle$, d-e) and spin ($\langle S_x\rangle$, g-h), resolved into Eu and SX contributions, is shown for comparison. In (j-l), the distribution of orbital currents arising in response to the exchange torque along $y$, as given by tensor components $t^{L_z}_{xy}$ and $t^{L_y}_{xy}$ defined by Eq.~\ref{t}, is shown.  In (m-o), the respective tensors are shown for the case of spin current. The band filling dependence of the orbital-to-orbital-current conversion strength, as given by tensors 
$\tau_{x}^{xz}$ and $\tau_{x}^{xy}$ defined by Eq.~\ref{J}, is shown for comparison in (p-r). 
The case of EuSP, EuSSe, and EuSCl corresponds to the left, middle, and right columns, respectively.}
    \label{fig:Torque}
\end{figure*}

\begin{figure*}[t!] \includegraphics[angle=270, width=0.85\textwidth]{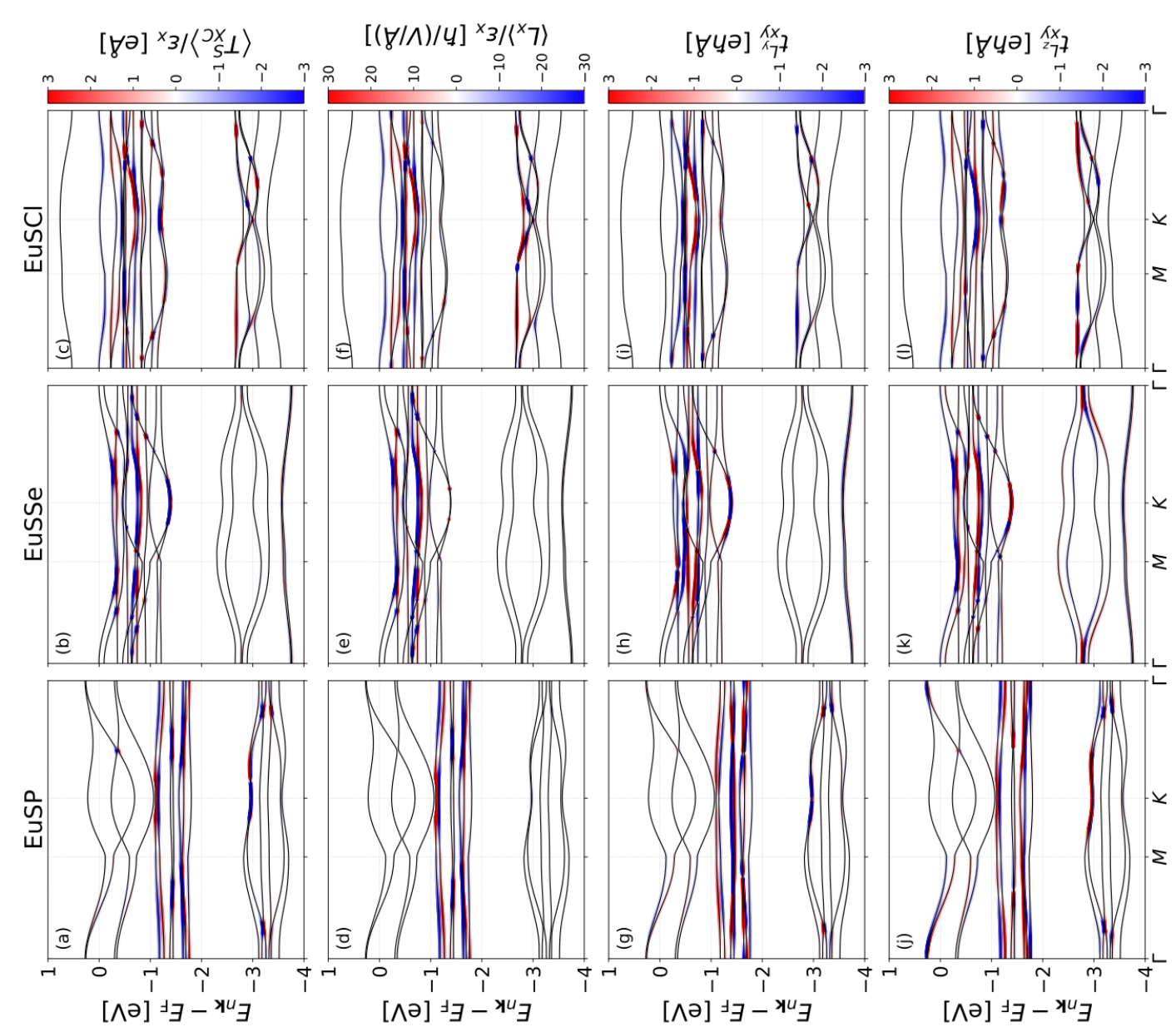}
    \caption{  {\bf Anatomy of orbital torques and orbital pumping in $k$-space.} For each compound, the $k$-resolved distribution is shown for (a-c) exchange torque $\langle T_{\rm XC}^\mathbf{S}\rangle$ normalized to the field $E_x$, (d-f) current induced orbital moment $\langle L_x\rangle$ normalized to the field $E_x$, and pumped by the torque along $y$ orbital current density along $x$ polarized along $y$ ($t^{L_y}_{xy}$, g-i), and along $z$ ($t^{L_z}_{xy}$, j-l). For all cases, the colored circles  represent the expectation value of the plotted quantity with the color code shown on the left.}
    \label{fig:Torque_band}
\end{figure*}

We present the results of our calculations for the $y$-component of the total exchange torque, i.e.~$t_{yx}$ component of the torkance tensor, and spin-orbital torque in Fig.~\ref{fig:Torque}(a-c) for EuSP, EuSSe, and EuSCl, respectively, as a function of the Fermi energy, keeping the magnetization aligned along $z$ in all cases. 
First,  we find that the overall magnitude of the exchange torque is very large, almost reaching $-2$\,e\AA, far exceeding that in transition metal heterostructures~\cite{freimuth2014spin}. This is comparable to the magnitude exhibited by some representative 2D materials~\cite{wang2019,xie2019,macneill2017,husain2020,hanke2017mixed}. For EuSCl, large values of the torque can be achieved in direct proximity of the Fermi level, which underlines the promise of this particular material for spin-orbit torque applications.  Expectedly, the total  summed over all atoms exchange torque in these systems is equivalent to the orbital torque, which is also apparent from the fact that it is very difficult to distinguish the curves for the orbital and exchange torques in the whole interval of energies, Fig.~\ref{fig:Torque}(a-c).

By relating the regions in energy, where the exchange torque is non-zero, to the band-resolved contributions shown in Fig.~\ref{fig:Torque_band}(a-c), we identify the largest torque with the region of $4f$-states, observing that a much smaller torque also arises in the region of $p$-states below the $f$-region (for EuSP).  Correspondingly, we identify Eu atoms as contributing predominantly to the orbital ``$f$"-torque, although at some energies within the $f$-band of EuSCl the contributions of S and Cl atoms are also non-zero (not shown), owing to the $p$-$d$-$f$ hybridization taking place in that region of energy. The origin of the torque in near band degeneracies becomes apparent after a careful analysis of band-resolved contributions: for example, while the two flat $p$-bands positioned at about $-3.2$\,eV in EuSP contribute only at the points where they cross the more dispersive bands, the latter provide a contribution over a larger part of the $\boldsymbol{k}$-path  as they are nearly degenerate around $-3$\,eV. The roles of these two groups of $p$-states are interchanged in EuSSe, with flat bands being nearly degenerate resulting in a small but finite torque, while the more dispersive bands are split off from each other almost everywhere giving a vanishing torque signal.

Next, to gain further insight into the origins of the torque, we analyze the non-equilibrium interband spin and orbital accumulation induced by an external electric field and calculated according to the linear response theory from the following expression:
\begin{eqnarray}
 \frac{\left\langle \mathbf{J} \right\rangle}{e\hbar E_x}  
= 2 \sum_{n\boldsymbol{k}<E_F}^{m\boldsymbol{k}>E_F} 
\operatorname{Im} \left[ \frac{\left \langle u_{n\boldsymbol{k}} | {\mathbf{{J}}}  | u_{m\boldsymbol{k}} \right \rangle \left \langle u_{m\boldsymbol{k}} | v_x | u_{n\boldsymbol{k}} \right \rangle}{{N_{\boldsymbol{k}}(E_{n\boldsymbol{k}} - E_{m\boldsymbol{k}}+i\eta )^2} }\right],
\nonumber 
\\
\end{eqnarray}
where $\mathbf{{J}}$ stands either for the operator of spin or OAM, $\mathbf{{S}}$ or $\mathbf{{L}}$. The results are presented in  Fig.~\ref{fig:Torque}(d-i) for the only non-vanishing components $\langle L_x\rangle$ and $\langle S_x\rangle$, decomposed into Eu and SX contributions.
The exchange $f$-torque observed in the monolayers of EuSX systems, as shown in Fig.~\ref{fig:Torque}, can be primarily attributed to orbital physics, which clearly manifests in a remarkable  one-to-one correspondence in the band filling behavior of  $\langle L_x\rangle$ on Eu atoms and $\langle T_{\rm XC}^\mathbf{S}\rangle$ in the region of $f$-states. 
This observation highlights the key role played by nonequilibrium OAM generation in mediating the overall torque in 2D limit. Note, that the $\langle L_x\rangle$ of SX atoms is non-vanishing only for the case of EuSCl, where it is still an order of magnitude smaller than that on Eu atoms. In particular, for EuSCl, the non-equilibrium orbital magnetism is of purely $p$-origin in the $p$-region of states below $-$3\,eV, see also Fig.~\ref{fig:Torque_band}(f). The impact of the $p$-states in the $f$-region on the Eu-$\langle L_x\rangle$ is also significant for EuSCl, as the behavior of $\langle L_x\rangle$ and the torque for this material is distinctly different from the other two materials $-$ something, which can be clearly observed also in Fig.~\ref{fig:Torque_band}(d-f).   

The analysis of current-induced spin polarization, $\langle S_x\rangle$, in Fig.~\ref{fig:Torque}(g-i) confirms the origin of the non-equilibrium spin density on Eu atoms in the current-induced orbital magnetization by SOC. Indeed,  although the magnitude of Eu $\langle S_x\rangle$ is about two orders of magnitude smaller than that of $\langle L_x\rangle$, the close correlation between the two quantities in the $f$-region is obvious.
At the same time, isolated regions, indicated by small black arrows (marked as A, B, C and D) in Fig.~\ref{fig:Torque}(h-i), where the correlation among Eu $\langle S_x\rangle$ and $\langle L_x\rangle$ is violated, are also present. These contributions to $\langle S_x\rangle$ can be traced back to hybridization points between Eu and SX atoms, which drive the redistribution of spin density among atoms. 

Remarkably, the latter points do not contribute to the exchange torque via $T_{\rm XC}^\mathbf{S}
= -(2\mu_{B}/\hbar){\mathbf{S}}\times\boldsymbol{V}_{\rm XC}$, which is expressed in terms of the operator of spin and exchange field~\cite{freimuth2014spin}. This effectively results in an observation that the current-induced orbital magnetization correlates with the exchange torque much stronger than the non-equilibrium spin density, despite the fact that the operator of OAM does not participate in $T_{\rm XC}^\mathbf{S}$ explicitly. This can be attributed to the fact that the overlap of the induced spin density $S_x(\mathbf{r})$ at chalcogen atoms S and X with strongly localized $\boldsymbol{V}_{\rm XC}(\mathbf{r})$ is small, with the former being inconsistent in symmetry with the angular expansion of the $f$-electron driven exchange field. On the other hand, the local orbital response $\langle L_x\rangle$ on Eu atoms is very localized as well, and it stems predominantly from the $f$-states which ensures the locality and suitable symmetry of the SOC-induced spin density. Our findings thus establish the local current-induced OAM as the key factor in mediating spin-orbit torques in $f$-systems.

\vspace{0.2cm}
\textbf{Orbital pumping in  EuSX}.
Next, we focus on the spin and orbital current generation by magnetization dynamics. To model the effect of the magnetization dynamics, we asumme that the exchange field becomes time-dependent via the time-dependence of the magnetization direction: 
\begin{equation}
\boldsymbol{V}_{\rm XC}(t)=\boldsymbol{\hat{M}}(t)\cdot V_{\rm XC}(\mathbf{r}).
\end{equation} 
It can be shown~\cite{freimuth2015direct} that this results in the following rate of change of the Hamiltonian of the system:
\begin{equation}
\frac{d H}{d t}=T_{\rm XC}^\mathbf{S}\cdot\left(\boldsymbol{\hat{M}}(t)\times\frac{d \boldsymbol{\hat{M}}(t)}{d t}\right).
\label{dhdt}
\end{equation}
In terms of the torkance tensor $t_{\alpha\beta}$, by using perturbation theory, one can arrive at the expression for the so-called inverse spin-orbit torque effect, i.e. an effect of charge pumping by magnetization dynamics:
\begin{equation}
j_\alpha(t)=\frac{1}{V}\sum_{\beta}t_{\beta\alpha}\left(\boldsymbol{\hat{M}}(t)\right)\left(\boldsymbol{\hat{M}}(t)\times\frac{d \boldsymbol{\hat{M}}(t)}{dt}\right)_\beta,
\end{equation}
where $V$ is the volume and $j_\alpha$ is the $\alpha$'th component of the generated by magnetization dynamics charge current density~\cite{freimuth2015direct}. Thus, following Onsager's reciprocity principle, the torkance tensor describes both the response of torque to the current, and the response of current to the magnetization dynamics.

In this context, one can also ask a question whether the time-dependent magnetization can give rise to the non-vanishing spin and orbital currents, arising in addition to the currents of charge. In the past, Kubo linear response formalism has been used to compute the effect of spin pumping $-$ i.e. spin current driven by magnetization dynamics $-$ into the substrate by a deposited thin ferromagnet,~e.g.~\cite{freimuth2015direct}, with the effect being extensively studied experimentally~\cite{ althammer2015spin}. In a similar geometry the twin effect of orbital pumping has been also recently observed~\cite{hayashi2024observation}. In contrast to the previous works where the currents of spin and OAM propagate into the bulk of the film, here, we consider a situation where the corresponding currents are lying in the plane of our EuSX systems, see Fig.~\ref{fig:pumping}(c), constituting thereby the effect of {\it in-plane orbital pumping}.

To assess the effects of orbital and spin pumping, we introduce the inverse spin and orbital torkance tensors $t^{J_\gamma}_{\beta\alpha}$ according to
\begin{equation}
Q^{J_\gamma}_\alpha(t)=\frac{1}{V}\sum_{\beta}t^{J_\gamma}_{\beta\alpha}\left(\boldsymbol{\hat{M}}(t)\right)\left(\boldsymbol{\hat{M}}(t)\times\frac{d \boldsymbol{\hat{M}}(t)}{d t}\right)_\beta,
\end{equation}
where $J_\gamma$ stands for the $\gamma$-component of either spin ($S_\gamma$) or orbital ($L_\gamma$) angular momentum, and $Q^{J_\gamma}_\alpha$ is the current density of $J_\gamma$ along direction $\alpha$. Within the Kubo linear response the intrinsic contribution to the inverse spin (orbital) torkance tensor reads: 
\begin{equation}\label{t}
 \frac{t^{J_\gamma}_{\alpha\beta}}{e\hbar}  
= 2 \sum_{n\boldsymbol{k}<E_F}^{m\boldsymbol{k}>E_F} 
\operatorname{Im} \left[ \frac{\left \langle u_{n\boldsymbol{k}} | j^{J_\gamma}_\alpha  | u_{m\boldsymbol{k}} \right \rangle \left \langle u_{m\boldsymbol{k}} | T_{\rm {XC},\beta}^\mathbf{S} | u_{n\boldsymbol{k}} \right \rangle}{N_{\boldsymbol{k}}(E_{n\boldsymbol{k}} - E_{m\boldsymbol{k}}+i\eta )^2}\right],
\end{equation}
where $j^{J_\gamma}_\alpha=\frac{1}{2}(v_{\alpha}{J}^{\gamma} + {J}^{\gamma}v_{\alpha})$ is the operator of spin (orbital) velocity.

For our compounds, the symmetry dictates that for the case of the torque along $y$ ($\beta=y$) the only non-vanishing components are $t_{xy}^{J_z}$, $t_{xy}^{J_y}$,   and $t_{yy}^{J_x}= t_{xy}^{J_y}$,
while for the torque along $x$ ($\beta=x$) the non-vanishing components are $t_{yx}^{J_z}=-t_{xy}^{J_z}$, $t_{yx}^{J_x}=-t_{xy}^{J_y}$, and $t_{xx}^{J_x}= -t_{yx}^{  J_y}$. 
In other words, when a torque is applied  on the magnetization, this results in a generation of two types of currents of angular momentum: \\(i) A current which is polarized along  the direction of the magnetization and propagates together with the pumped flow of charge; 
\\
(ii) An in-plane current of angular momentum with an in-plane propagation-direction-dependent polarization, whose properties can be controlled by the direction of the applied torque. 
\\
It is important to realize that were our atoms arranged in a square lattice with both mirror symmetries $\mathcal{M}_x$ and $\mathcal{M}_y$ present, only one, conventional, component  $t_{xy}^{J_z}$ would survive, driving the current of type (i). The complexity of the pumped currents of angular momentum is thus a direct consequence of the complex structural symmetry of  considered Janus rare-earth dichacogenides.

The  tensor components $t_{xy}^{S_z}$ and $t_{xy}^{S_y}$ for the case of spin, and $t_{xy}^{L_z}$ and $t_{xy}^{L_y}$ for the case of  orbital channel are shown for EuSX in Fig.~\ref{fig:Torque}(j-l) and Fig.~~\ref{fig:Torque}(m-o), respectively, as a function of band filling. 
By first looking at the pumped spin currents, we immediately realize that for all compounds, in case of the $z$-polarization, the spin current in the $f$-region is directly proportional to the generated by $E_x$ distribution of $\langle T_{\rm XC}^\mathbf{S}\rangle$ and $\langle L_x\rangle$, and not to the corresponding values of $\langle S_x\rangle$, Fig.~~\ref{fig:Torque}(a-i). This is rooted in the fact that in the ground state the localized $f$-states, on which the torque predominantly operates, are almost pure eigenstates of $\langle S_z\rangle$, which results in proportionality of $Q_x^{S_z}$ to $\langle T_{\rm XC}^\mathbf{S}\rangle$. On the other hand, for similar reasons, the $Q_y^{S_y}$ is vanishing almost everywhere except for the spin-flip points where the spin-up and spin-down bands of SX and Eu atoms meet (observe for example vanishing $t_{xy}^{S_y}$ for EuSP where $f$-states are well-separated from SP-states). 

The behavior of the pumped orbital currents is, however, very different.
We observe that, qualitatively, in the $f$-region the $z$-polarized orbital current 
can be reconstructed from the behavior of the $\langle T_{\rm XC}^\mathbf{S}\rangle$ and $\langle L_x\rangle$ in the lower (negative $\langle L_z\rangle$) and upper (positive $\langle L_z\rangle$) $f$-band, qualitatively multiplied with the sign of $\langle L_z\rangle$ (not shown), while being suppressed in the middle of the $f$-band, where the transition from positive to negative $\langle L_z\rangle$ occurs. In an effective picture, one can thus preceive pumped orbital current $Q_x^{L_z}$ as an orbital filtering effect, where the pumped charge current gets orbitally polarized along $z$ as it is generated by an application of a torque. This picture has to be taken with caution, however, since it is the off-diagonal matrix elements of OAM operator which ultimately participate in the Eq.~\ref{t}, and which do not necessarily reflect the orbital polarization of the states in the ground state. 

When inspecting the in-plane polarized OAM current, we find that it is peaking in the middle of the $f$-band, where $z$-polarized current is suppressed. In contrast to $Q_x^{S_y}$,  $Q_x^{L_y}$ is spread over wider regions of energy, owing to a higher complexity in the orbital composition of the bands which allows for richer coupling among the states with different values of OAM. On the other hand, when looking at the $k$-space behavior of the two types of orbital currents, shown in Fig.~\ref{fig:Torque_band}(g-l), we find two clear distinctions: first, $Q_x^{L_y}$ is originating from narrower regions in the reciprocal space, and second, $Q_x^{L_z}$ exhibits variations in sign when going along a given band, while $Q_x^{L_z}$ is much more uniform when the $k$-point is varied within the regions of strong contributions.

Above, we have observed that for $f$-electrons the effect of the current-induced torque on the exchange field is equivalent to the effect of current-induced orbital magnetization. Therefore, it seems rewarding to explore whether the same holds true for the inverse process:~i.e., whether the pumping of the OAM currents by the torque is equivalent to an OAM current arising in response  to an induced orbital magnetization.
Physically, we have to assess the response of the system in terms of a  pumped  OAM current  when the orbital magnetization  is increasing in time $-$ an effect which can be modelled by considering an additional orbital term in the Hamiltonian,  $B_L(t)\mathbf{L}$, with linearly increasing in time orbital exchange field, $B_L(t)\sim t$. Introducing such a term on the level of spin  has been used in the past to study the phenomena of inverse (spin) Edelstein effect~\cite{shen2014microscopic} and its magnetic counterpart~\cite{freimuth2017}, associated with the pumping of charge currents. By considering this term as a perturbation, we are interested in the response of the orbital currents, proportional to the rate of change in the orbital magnetic field,  which is governed by the following {\it orbital-to-orbital-current  conversion} tensor:
\begin{equation}\label{J}
 \frac{\tau_{\alpha}^{\beta\gamma}}{e\hbar}  
= 2 \sum_{n\boldsymbol{k}<E_F}^{m\boldsymbol{k}>E_F} 
\operatorname{Im} \left[ \frac{\left \langle u_{n\boldsymbol{k}} | j^{L_\gamma}_\beta  | u_{m\boldsymbol{k}} \right \rangle \left \langle u_{m\boldsymbol{k}} | \hat{L}_{\alpha} | u_{n\boldsymbol{k}} \right \rangle}{N_{\boldsymbol{k}}(E_{n\boldsymbol{k}} - E_{m\boldsymbol{k}}+i\eta )^2}\right].
\end{equation}
The calculations for the $\tau_x^{xz}$ and $\tau_x^{xy}$ components $-$ corresponding to a generation of $y$- or $z$-polarized OAM currents along $x$ in response to the orbital field applied along $x$ $-$ are shown in Fig.~\ref{fig:Torque}(p-r). From this data we make a remarkable observation: as far as pumped orbital currents in the region of the $f$-states are concerned, the effect of the magnetization dynamics represented by a torque along $y$ is {\it quantitatively} completely equivalent to the effect of $L_x$ via the application of a time-dependent orbital field, since the two types of tensors exhibit an almost identical to each other behavior in the region of corresponding energies. 
Conceptually, this is similar to the picture of the origin of the inverse spin Edelstein effect in a linearly increasing magnetic field acting on spin~\cite{shen2014microscopic}. The effect of the OAM current generation by inverse orbital torque that we uncover here can be correspondingly qualitatively considered as an intrinsic version of the inverse orbital Rashba effect for the orbital current, see Fig.~\ref{fig:pumping}.

\begin{table*}[t!]

\caption{Optimal lattice parameters and magnetic properties of EuSX for out of plane magnetization: lattice constant (a), buckled height between Eu-S/X atoms plane $\Delta$(Eu-S/X), and the magnetic moment (S) of atom Eu and S/X, as well as the orbital magnetic moment (L) of Eu, in addition to the bandgap.}

\renewcommand{\arraystretch}{1.2}

\begin{tabular}{p{0.06\textwidth}
>{\centering}p{0.105\textwidth}
>{\centering}p{0.105\textwidth}
>{\centering}p{0.105\textwidth}
>{\centering}p{0.105\textwidth}
>{\centering}p{0.105\textwidth}
>{\centering}p{0.105\textwidth}
>{\centering}p{0.105\textwidth}
>{\centering\arraybackslash}p{0.11\textwidth}}
\hline
\hline
&Lattice constant [\AA]
&$d_\mathrm{Eu-S}$\ \ \ \ \ \ \ \ [\AA]
&$d_\mathrm{Eu-S}$\ \ \ \ \ \ \ \ [\AA]
& $m_S$ (Eu) \ \ \ [$\mu_B$]
& $m_S$ (S) \ \ \ \ \ [$\mu_B$]
& $m_S$ (X) \ \ \ \ \ [$\mu_B$]
& $m_L$ (Eu) \ \ \ [$\mu_B$]
& Band gap \ \ \ [eV]\\
\hline
EuSP& 4.789&1.136&1.008&6.829&$-$0.337&$-$0.120&$-$0.047&0.002\\
EuSSe&4.785&1.029&1.307&6.951&$-$0.022&$-$0.024&$-$0.007&1.847\\
EuSCl&4.479&1.315&1.571&6.875&$-$0.638&$-$0.192&$-$0.035&0.534\\
\hline
\hline
\label{table:Lattice}
\end{tabular}
\end{table*}%

\section*{CONCLUDING REMARKS}

Our study provides unique insights into the structural, electronic, magnetic, and transport properties of Janus H-phase monolayers of 4$f$-Eu rare-earth dichalcogenides. The observation of substantial magnetic moments, distinct electronic hybridization channels, and non-trivial response to an applied current highlights the potential of these materials for spintronic device applications. The understanding gained from this research opens new avenues for designing and optimizing orbitronics devices based on 4$f$-Eu rare-earth dichalcogenides, thus contributing to the broader knowledge of electronic and transport properties of 2D magnetic materials.

One of our key findings is  the observation of pronouncedly orbital character of the response to the electrical current which is mediated by $f$-electrons. The close correlation between the orbital magnetism and torques in $f$-systems suggest that the orbital torque properties can be engineered by carefully crafting the crystal field around rare-earth atoms due to surrounding cage of $sp$-atoms. The latter is susceptible to strong changes in 2D geometry which signifies a powerful channel for spin-orbit torque design. The found correlation between the orbital magnetism and torque within the $f$-shell is so strong that the torque operator acting on $f$-states can be replaced by the local OAM operator within the same sub-shell in the linear response expressions for the response properties. We attribute this to the properties of the exchange field of $f$-atoms, which is closely interlinked with the strong orbital polarization of the $f$-states by very large spin-orbit interaction. Since the properties of the $f$-states, their orbital polarization and degree of localization are very sensitive to the degree of correlation of valence electrons, one possible way to tune the response properties in considered classes of systems lies in alloying with other rare-earth or transition-metals.   

Another remarkable property  that we uncovered is the effect of in-plane orbital pumping, i.e. pumping of currents of OAM by magnetization dynamics. Intuitively, one would expect that this effect can be attributed to the OAM polarized along the magnetization direction and carried by the pumped current of charge. However, we find  this picture to be too naive to reflect the complexity that we observe in the magnitude, direction and polarization of the generated current. We speculate that this complexity may be exploited for the purpose of orbital current injection into a planar adjacent material, with a controllable polarization of the current. We predict  that this is a generic effect which occurs in 2D magnetic materials of Janus type owing to their lowered crystal symmetry. While we leave the exploration of exact microscopic mechanisms for the orbital pumping to future work, we make the first step in this direction by discovering that orbital pumping realized by $f$-electrons can be considered as an effective orbital-to-orbital-current conversion, where changing in time orbital polarization relaxes via a generation of an orbital current. Our findings open new possibilities for generating diverse flavors of orbital currents in 2D materials which go beyond the protocol of the OHE. We thus declare that the peculiarities of 2D geometry combined with the specifics of $f$-electron magnetism may provide as robust and efficient platform for orbital current generation.   

\section*{METHODS}
We performed our density functional theory (DFT) calculations of Janus H-phase monolayers of EuSX (X = P, Se, Cl) using the full-potential linearized augmented plane wave method (FLAPW), as implemented in the Jülich DFT code FLEUR \cite{fleur}, with the Perdew-Burke-Ernzerhof \cite{perdew1996generalized} parametrization of the exchange-correlation potential. The structural optimization  was carried out using the FLEUR code with DFT+$U$ method. For the self-consistent calculations, a 16$\times$16 $k$-point mesh was used to sample the first Brillouin zone. All calculations included SOC self-consistently within the second variation scheme. To account for the effect of strongly correlated electrons of the highly localized 4$f$-electrons of Eu, the GGA+$U$ method was applied within a self-consistent DFT cycle. The on-site Coulomb and Hund exchange parameters were set to $U$ = 6.7 eV and $J$ = 0.7 eV, which are values commonly used for the treatment of $f$ electrons in chemical elements with half-filled 4f shells such as Eu and Gd \cite{Shick1999,kurz2002}.

Table \ref{table:Lattice} summarizes the relaxed 
structure lattice parameters, 
magnetic moments of Eu, S, and X atoms, and the band gaps calculated using the GGA+$U$ method. 
Regarding the computational parameters, we set the muffin-tin radii to 2.80 $a_0$ for Eu and 1.69, 1.84 and 2.35  $a_0$ for S, and 2.16, 2.35 and 2.95  $a_0$ for X, where  $a_0$ is the Bohr radius. The cut-off for the plane-wave basis functions was chosen as $K_{\text{max}}$ = 3.7, 4.3 and 3.9\,$a_0 ^{-1}$ for the wave functions and $G_{\text{max}}$ = 11.1, 12.9 and 11.7\,$a_0 ^{-1}$ for the charge density and potential for EuSP, EuSSe and EuSCl, respectively. The upper limit of angular momentum inside of the muffin-tin sphere was set to $l_{\text{max}}$ = 10 for Eu and $l_{\text{max}}$ = 6, 8 and 10  for S and 8, 8 and 10 for X. The magnetic anisotropy energy of EuSP, EuSSe and EuSCl was estimated to be 0.73,  0.047 and 2.341 meV, respectively, favoring an out-of-plane direction of magnetization.

From DFT calculations, we extracted 36 maximally localized Wannier functions (MLWFs) 
by using Eu-$d$, Eu-$f$ and S,X-$p$ orbitals as initial projections, as these orbitals dominate the band structure in a wide energy window around the Fermi energy, with the frozen window maximum set to 2 eV above the Fermi energy. The electronic band structure was compared to the band structure obtained by constructing MLWFs in the FLAPW formalism \cite{Freimuth2008} and the open-source WANNIER90 \cite{MOSTOFI20142309}. 
Based on the tight-binding Wannier Hamiltonian constructed from the Wannier functions, we performed the calculations on a 500$\times$500 interpolation k-mesh. The orbital pumping and orbital torque were computed using the Kubo formula and the Wannier interpolation technique, capturing the dominant contributions from the specified orbitals.

\begin{acknowledgements}
The authors appreciate fruitful discussions with Dr. Tom G. Saunderson.
This work was supported by the Federal Ministry of Education and Research of Germany in the framework of the Palestinian-German Science Bridge (BMBF grant number 01DH16027). We also gratefully acknowledge financial support by the Deutsche Forschungsgemeinschaft (DFG, German Research Foundation) $-$ TRR 288 $-$ 422213477 (project B06),  TRR 173/2 $-$ 268565370 (projects A11 and A01), CRC 1238 - 277146847 (Project C01), and the Sino-German research project DISTOMAT (MO 1731/10-1).  This work was supported by the EIC Pathfinder OPEN grant 101129641 “OBELIX”. We  also gratefully acknowledge the J\"ulich Supercomputing Centre and RWTH Aachen University for providing computational resources under projects jiff40 and jara0062. 

\end{acknowledgements}

\bibliography{bib_EuSCl_EuSSe_EuSP}

\begin{thebibliography}{75}%
\makeatletter
\providecommand \@ifxundefined [1]{%
 \@ifx{#1\undefined}
}%
\providecommand \@ifnum [1]{%
 \ifnum #1\expandafter \@firstoftwo
 \else \expandafter \@secondoftwo
 \fi
}%
\providecommand \@ifx [1]{%
 \ifx #1\expandafter \@firstoftwo
 \else \expandafter \@secondoftwo
 \fi
}%
\providecommand \natexlab [1]{#1}%
\providecommand \enquote  [1]{``#1''}%
\providecommand \bibnamefont  [1]{#1}%
\providecommand \bibfnamefont [1]{#1}%
\providecommand \citenamefont [1]{#1}%
\providecommand \href@noop [0]{\@secondoftwo}%
\providecommand \href [0]{\begingroup \@sanitize@url \@href}%
\providecommand \@href[1]{\@@startlink{#1}\@@href}%
\providecommand \@@href[1]{\endgroup#1\@@endlink}%
\providecommand \@sanitize@url [0]{\catcode `\\12\catcode `\$12\catcode
  `\&12\catcode `\#12\catcode `\^12\catcode `\_12\catcode `\%12\relax}%
\providecommand \@@startlink[1]{}%
\providecommand \@@endlink[0]{}%
\providecommand \url  [0]{\begingroup\@sanitize@url \@url }%
\providecommand \@url [1]{\endgroup\@href {#1}{\urlprefix }}%
\providecommand \urlprefix  [0]{URL }%
\providecommand \Eprint [0]{\href }%
\providecommand \doibase [0]{https://doi.org/}%
\providecommand \selectlanguage [0]{\@gobble}%
\providecommand \bibinfo  [0]{\@secondoftwo}%
\providecommand \bibfield  [0]{\@secondoftwo}%
\providecommand \translation [1]{[#1]}%
\providecommand \BibitemOpen [0]{}%
\providecommand \bibitemStop [0]{}%
\providecommand \bibitemNoStop [0]{.\EOS\space}%
\providecommand \EOS [0]{\spacefactor3000\relax}%
\providecommand \BibitemShut  [1]{\csname bibitem#1\endcsname}%
\let\auto@bib@innerbib\@empty
\bibitem [{\citenamefont {Shao}\ \emph {et~al.}(2021)\citenamefont {Shao},
  \citenamefont {Li}, \citenamefont {Liu}, \citenamefont {Yang}, \citenamefont
  {Fukami}, \citenamefont {Razavi}, \citenamefont {Wu}, \citenamefont {Wang},
  \citenamefont {Freimuth}, \citenamefont {Mokrousov} \emph
  {et~al.}}]{shao2021roadmap}%
  \BibitemOpen
  \bibfield  {author} {\bibinfo {author} {\bibfnamefont {Q.}~\bibnamefont
  {Shao}}, \bibinfo {author} {\bibfnamefont {P.}~\bibnamefont {Li}}, \bibinfo
  {author} {\bibfnamefont {L.}~\bibnamefont {Liu}}, \bibinfo {author}
  {\bibfnamefont {H.}~\bibnamefont {Yang}}, \bibinfo {author} {\bibfnamefont
  {S.}~\bibnamefont {Fukami}}, \bibinfo {author} {\bibfnamefont
  {A.}~\bibnamefont {Razavi}}, \bibinfo {author} {\bibfnamefont
  {H.}~\bibnamefont {Wu}}, \bibinfo {author} {\bibfnamefont {K.}~\bibnamefont
  {Wang}}, \bibinfo {author} {\bibfnamefont {F.}~\bibnamefont {Freimuth}},
  \bibinfo {author} {\bibfnamefont {Y.}~\bibnamefont {Mokrousov}}, \emph
  {et~al.},\ }\bibfield  {title} {\bibinfo {title} {Roadmap of spin--orbit
  torques},\ }\href@noop {} {\bibfield  {journal} {\bibinfo  {journal} {IEEE
  transactions on magnetics}\ }\textbf {\bibinfo {volume} {57}},\ \bibinfo
  {pages} {1} (\bibinfo {year} {2021})}\BibitemShut {NoStop}%
\bibitem [{\citenamefont {Jo}\ \emph {et~al.}(2024)\citenamefont {Jo},
  \citenamefont {Go}, \citenamefont {Choi},\ and\ \citenamefont
  {Lee}}]{jo2024spintronics}%
  \BibitemOpen
  \bibfield  {author} {\bibinfo {author} {\bibfnamefont {D.}~\bibnamefont
  {Jo}}, \bibinfo {author} {\bibfnamefont {D.}~\bibnamefont {Go}}, \bibinfo
  {author} {\bibfnamefont {G.-M.}\ \bibnamefont {Choi}},\ and\ \bibinfo
  {author} {\bibfnamefont {H.-W.}\ \bibnamefont {Lee}},\ }\bibfield  {title}
  {\bibinfo {title} {Spintronics meets orbitronics: Emergence of orbital
  angular momentum in solids},\ }\href@noop {} {\bibfield  {journal} {\bibinfo
  {journal} {npj Spintronics}\ }\textbf {\bibinfo {volume} {2}},\ \bibinfo
  {pages} {19} (\bibinfo {year} {2024})}\BibitemShut {NoStop}%
\bibitem [{\citenamefont {Wang}\ \emph {et~al.}(2024)\citenamefont {Wang},
  \citenamefont {Chen}, \citenamefont {Yang}, \citenamefont {Hu}, \citenamefont
  {Li}, \citenamefont {Wang}, \citenamefont {Zhang},\ and\ \citenamefont
  {Jiang}}]{wang2024orbitronics}%
  \BibitemOpen
  \bibfield  {author} {\bibinfo {author} {\bibfnamefont {P.}~\bibnamefont
  {Wang}}, \bibinfo {author} {\bibfnamefont {F.}~\bibnamefont {Chen}}, \bibinfo
  {author} {\bibfnamefont {Y.}~\bibnamefont {Yang}}, \bibinfo {author}
  {\bibfnamefont {S.}~\bibnamefont {Hu}}, \bibinfo {author} {\bibfnamefont
  {Y.}~\bibnamefont {Li}}, \bibinfo {author} {\bibfnamefont {W.}~\bibnamefont
  {Wang}}, \bibinfo {author} {\bibfnamefont {D.}~\bibnamefont {Zhang}},\ and\
  \bibinfo {author} {\bibfnamefont {Y.}~\bibnamefont {Jiang}},\ }\bibfield
  {title} {\bibinfo {title} {Orbitronics: Mechanisms, materials and devices},\
  }\href@noop {} {\bibfield  {journal} {\bibinfo  {journal} {Advanced
  Electronic Materials}\ ,\ \bibinfo {pages} {2400554}} (\bibinfo {year}
  {2024})}\BibitemShut {NoStop}%
\bibitem [{\citenamefont {Xu}\ \emph {et~al.}(2024)\citenamefont {Xu},
  \citenamefont {Zhang}, \citenamefont {Fert}, \citenamefont {Jaffres},
  \citenamefont {Liu}, \citenamefont {Xu}, \citenamefont {Jiang}, \citenamefont
  {Cheng},\ and\ \citenamefont {Zhao}}]{xu2024orbitronics}%
  \BibitemOpen
  \bibfield  {author} {\bibinfo {author} {\bibfnamefont {Y.}~\bibnamefont
  {Xu}}, \bibinfo {author} {\bibfnamefont {F.}~\bibnamefont {Zhang}}, \bibinfo
  {author} {\bibfnamefont {A.}~\bibnamefont {Fert}}, \bibinfo {author}
  {\bibfnamefont {H.-Y.}\ \bibnamefont {Jaffres}}, \bibinfo {author}
  {\bibfnamefont {Y.}~\bibnamefont {Liu}}, \bibinfo {author} {\bibfnamefont
  {R.}~\bibnamefont {Xu}}, \bibinfo {author} {\bibfnamefont {Y.}~\bibnamefont
  {Jiang}}, \bibinfo {author} {\bibfnamefont {H.}~\bibnamefont {Cheng}},\ and\
  \bibinfo {author} {\bibfnamefont {W.}~\bibnamefont {Zhao}},\ }\bibfield
  {title} {\bibinfo {title} {Orbitronics: light-induced orbital currents in ni
  studied by terahertz emission experiments},\ }\href@noop {} {\bibfield
  {journal} {\bibinfo  {journal} {Nature Communications}\ }\textbf {\bibinfo
  {volume} {15}},\ \bibinfo {pages} {2043} (\bibinfo {year}
  {2024})}\BibitemShut {NoStop}%
\bibitem [{\citenamefont {Tanaka}\ \emph {et~al.}(2008)\citenamefont {Tanaka},
  \citenamefont {Kontani}, \citenamefont {Naito}, \citenamefont {Naito},
  \citenamefont {Hirashima}, \citenamefont {Yamada},\ and\ \citenamefont
  {Inoue}}]{Tanaka2008}%
  \BibitemOpen
  \bibfield  {author} {\bibinfo {author} {\bibfnamefont {T.}~\bibnamefont
  {Tanaka}}, \bibinfo {author} {\bibfnamefont {H.}~\bibnamefont {Kontani}},
  \bibinfo {author} {\bibfnamefont {M.}~\bibnamefont {Naito}}, \bibinfo
  {author} {\bibfnamefont {T.}~\bibnamefont {Naito}}, \bibinfo {author}
  {\bibfnamefont {D.~S.}\ \bibnamefont {Hirashima}}, \bibinfo {author}
  {\bibfnamefont {K.}~\bibnamefont {Yamada}},\ and\ \bibinfo {author}
  {\bibfnamefont {J.}~\bibnamefont {Inoue}},\ }\bibfield  {title} {\bibinfo
  {title} {{Intrinsic spin Hall effect and orbital Hall effect in $4d$ and $5d$
  transition metals}},\ }\href {https://doi.org/10.1103/PhysRevB.77.165117}
  {\bibfield  {journal} {\bibinfo  {journal} {Phys. Rev. B}\ }\textbf {\bibinfo
  {volume} {77}},\ \bibinfo {pages} {165117} (\bibinfo {year}
  {2008})}\BibitemShut {NoStop}%
\bibitem [{\citenamefont {Go}\ \emph {et~al.}(2018)\citenamefont {Go},
  \citenamefont {Jo}, \citenamefont {Kim},\ and\ \citenamefont {Lee}}]{Go2018}%
  \BibitemOpen
  \bibfield  {author} {\bibinfo {author} {\bibfnamefont {D.}~\bibnamefont
  {Go}}, \bibinfo {author} {\bibfnamefont {D.}~\bibnamefont {Jo}}, \bibinfo
  {author} {\bibfnamefont {C.}~\bibnamefont {Kim}},\ and\ \bibinfo {author}
  {\bibfnamefont {H.-W.}\ \bibnamefont {Lee}},\ }\bibfield  {title} {\bibinfo
  {title} {{Intrinsic Spin and Orbital Hall Effects from Orbital Texture}},\
  }\href {https://doi.org/10.1103/PhysRevLett.121.086602} {\bibfield  {journal}
  {\bibinfo  {journal} {Phys. Rev. Lett.}\ }\textbf {\bibinfo {volume} {121}},\
  \bibinfo {pages} {086602} (\bibinfo {year} {2018})}\BibitemShut {NoStop}%
\bibitem [{\citenamefont {Go}\ and\ \citenamefont {Lee}(2020)}]{go2020orbital}%
  \BibitemOpen
  \bibfield  {author} {\bibinfo {author} {\bibfnamefont {D.}~\bibnamefont
  {Go}}\ and\ \bibinfo {author} {\bibfnamefont {H.-W.}\ \bibnamefont {Lee}},\
  }\bibfield  {title} {\bibinfo {title} {{Orbital torque: Torque generation by
  orbital current injection}},\ }\href@noop {} {\bibfield  {journal} {\bibinfo
  {journal} {Physical review research}\ }\textbf {\bibinfo {volume} {2}},\
  \bibinfo {pages} {013177} (\bibinfo {year} {2020})}\BibitemShut {NoStop}%
\bibitem [{\citenamefont {Hayashi}\ \emph {et~al.}(2023)\citenamefont
  {Hayashi}, \citenamefont {Jo}, \citenamefont {Go}, \citenamefont {Gao},
  \citenamefont {Haku}, \citenamefont {Mokrousov}, \citenamefont {Lee},\ and\
  \citenamefont {Ando}}]{hayashi2023observation}%
  \BibitemOpen
  \bibfield  {author} {\bibinfo {author} {\bibfnamefont {H.}~\bibnamefont
  {Hayashi}}, \bibinfo {author} {\bibfnamefont {D.}~\bibnamefont {Jo}},
  \bibinfo {author} {\bibfnamefont {D.}~\bibnamefont {Go}}, \bibinfo {author}
  {\bibfnamefont {T.}~\bibnamefont {Gao}}, \bibinfo {author} {\bibfnamefont
  {S.}~\bibnamefont {Haku}}, \bibinfo {author} {\bibfnamefont {Y.}~\bibnamefont
  {Mokrousov}}, \bibinfo {author} {\bibfnamefont {H.-W.}\ \bibnamefont {Lee}},\
  and\ \bibinfo {author} {\bibfnamefont {K.}~\bibnamefont {Ando}},\ }\bibfield
  {title} {\bibinfo {title} {{Observation of long-range orbital transport and
  giant orbital torque}},\ }\href@noop {} {\bibfield  {journal} {\bibinfo
  {journal} {Communications Physics}\ }\textbf {\bibinfo {volume} {6}},\
  \bibinfo {pages} {32} (\bibinfo {year} {2023})}\BibitemShut {NoStop}%
\bibitem [{\citenamefont {Go}\ \emph {et~al.}(2023{\natexlab{a}})\citenamefont
  {Go}, \citenamefont {Jo}, \citenamefont {Kim}, \citenamefont {Lee},
  \citenamefont {Kang}, \citenamefont {Park}, \citenamefont {Bl{\"u}gel},
  \citenamefont {Lee},\ and\ \citenamefont {Mokrousov}}]{go2023long}%
  \BibitemOpen
  \bibfield  {author} {\bibinfo {author} {\bibfnamefont {D.}~\bibnamefont
  {Go}}, \bibinfo {author} {\bibfnamefont {D.}~\bibnamefont {Jo}}, \bibinfo
  {author} {\bibfnamefont {K.-W.}\ \bibnamefont {Kim}}, \bibinfo {author}
  {\bibfnamefont {S.}~\bibnamefont {Lee}}, \bibinfo {author} {\bibfnamefont
  {M.-G.}\ \bibnamefont {Kang}}, \bibinfo {author} {\bibfnamefont {B.-G.}\
  \bibnamefont {Park}}, \bibinfo {author} {\bibfnamefont {S.}~\bibnamefont
  {Bl{\"u}gel}}, \bibinfo {author} {\bibfnamefont {H.-W.}\ \bibnamefont
  {Lee}},\ and\ \bibinfo {author} {\bibfnamefont {Y.}~\bibnamefont
  {Mokrousov}},\ }\bibfield  {title} {\bibinfo {title} {{Long-range orbital
  torque by momentum-space hotspots}},\ }\href@noop {} {\bibfield  {journal}
  {\bibinfo  {journal} {Physical review letters}\ }\textbf {\bibinfo {volume}
  {130}},\ \bibinfo {pages} {246701} (\bibinfo {year}
  {2023}{\natexlab{a}})}\BibitemShut {NoStop}%
\bibitem [{\citenamefont {Lee}\ \emph {et~al.}(2021{\natexlab{a}})\citenamefont
  {Lee}, \citenamefont {Go}, \citenamefont {Park}, \citenamefont {Jeong},
  \citenamefont {Ko}, \citenamefont {Yun}, \citenamefont {Jo}, \citenamefont
  {Lee}, \citenamefont {Go}, \citenamefont {Oh} \emph
  {et~al.}}]{lee2021orbital}%
  \BibitemOpen
  \bibfield  {author} {\bibinfo {author} {\bibfnamefont {D.}~\bibnamefont
  {Lee}}, \bibinfo {author} {\bibfnamefont {D.}~\bibnamefont {Go}}, \bibinfo
  {author} {\bibfnamefont {H.-J.}\ \bibnamefont {Park}}, \bibinfo {author}
  {\bibfnamefont {W.}~\bibnamefont {Jeong}}, \bibinfo {author} {\bibfnamefont
  {H.-W.}\ \bibnamefont {Ko}}, \bibinfo {author} {\bibfnamefont
  {D.}~\bibnamefont {Yun}}, \bibinfo {author} {\bibfnamefont {D.}~\bibnamefont
  {Jo}}, \bibinfo {author} {\bibfnamefont {S.}~\bibnamefont {Lee}}, \bibinfo
  {author} {\bibfnamefont {G.}~\bibnamefont {Go}}, \bibinfo {author}
  {\bibfnamefont {J.~H.}\ \bibnamefont {Oh}}, \emph {et~al.},\ }\bibfield
  {title} {\bibinfo {title} {{Orbital torque in magnetic bilayers}},\
  }\href@noop {} {\bibfield  {journal} {\bibinfo  {journal} {Nature
  communications}\ }\textbf {\bibinfo {volume} {12}},\ \bibinfo {pages} {6710}
  (\bibinfo {year} {2021}{\natexlab{a}})}\BibitemShut {NoStop}%
\bibitem [{\citenamefont {Gambardella}\ and\ \citenamefont
  {Miron}(2011)}]{gambardella2011current}%
  \BibitemOpen
  \bibfield  {author} {\bibinfo {author} {\bibfnamefont {P.}~\bibnamefont
  {Gambardella}}\ and\ \bibinfo {author} {\bibfnamefont {I.~M.}\ \bibnamefont
  {Miron}},\ }\bibfield  {title} {\bibinfo {title} {Current-induced spin--orbit
  torques},\ }\href@noop {} {\bibfield  {journal} {\bibinfo  {journal}
  {Philosophical Transactions of the Royal Society A: Mathematical, Physical
  and Engineering Sciences}\ }\textbf {\bibinfo {volume} {369}},\ \bibinfo
  {pages} {3175} (\bibinfo {year} {2011})}\BibitemShut {NoStop}%
\bibitem [{\citenamefont {Dutta}\ and\ \citenamefont
  {Tulapurkar}(2022)}]{dutta2022observation}%
  \BibitemOpen
  \bibfield  {author} {\bibinfo {author} {\bibfnamefont {S.}~\bibnamefont
  {Dutta}}\ and\ \bibinfo {author} {\bibfnamefont {A.~A.}\ \bibnamefont
  {Tulapurkar}},\ }\bibfield  {title} {\bibinfo {title} {{Observation of
  nonlocal orbital transport and sign reversal of dampinglike torque in Nb/Ni
  and Ta/Ni bilayers}},\ }\href@noop {} {\bibfield  {journal} {\bibinfo
  {journal} {Physical Review B}\ }\textbf {\bibinfo {volume} {106}},\ \bibinfo
  {pages} {184406} (\bibinfo {year} {2022})}\BibitemShut {NoStop}%
\bibitem [{\citenamefont {Lee}\ \emph {et~al.}(2021{\natexlab{b}})\citenamefont
  {Lee}, \citenamefont {Kang}, \citenamefont {Go}, \citenamefont {Kim},
  \citenamefont {Kang}, \citenamefont {Lee}, \citenamefont {Lee}, \citenamefont
  {Kang}, \citenamefont {Lee}, \citenamefont {Mokrousov} \emph
  {et~al.}}]{lee2021efficient}%
  \BibitemOpen
  \bibfield  {author} {\bibinfo {author} {\bibfnamefont {S.}~\bibnamefont
  {Lee}}, \bibinfo {author} {\bibfnamefont {M.-G.}\ \bibnamefont {Kang}},
  \bibinfo {author} {\bibfnamefont {D.}~\bibnamefont {Go}}, \bibinfo {author}
  {\bibfnamefont {D.}~\bibnamefont {Kim}}, \bibinfo {author} {\bibfnamefont
  {J.-H.}\ \bibnamefont {Kang}}, \bibinfo {author} {\bibfnamefont
  {T.}~\bibnamefont {Lee}}, \bibinfo {author} {\bibfnamefont {G.-H.}\
  \bibnamefont {Lee}}, \bibinfo {author} {\bibfnamefont {J.}~\bibnamefont
  {Kang}}, \bibinfo {author} {\bibfnamefont {N.~J.}\ \bibnamefont {Lee}},
  \bibinfo {author} {\bibfnamefont {Y.}~\bibnamefont {Mokrousov}}, \emph
  {et~al.},\ }\bibfield  {title} {\bibinfo {title} {{Efficient conversion of
  orbital Hall current to spin current for spin-orbit torque switching}},\
  }\href@noop {} {\bibfield  {journal} {\bibinfo  {journal} {Communications
  Physics}\ }\textbf {\bibinfo {volume} {4}},\ \bibinfo {pages} {234} (\bibinfo
  {year} {2021}{\natexlab{b}})}\BibitemShut {NoStop}%
\bibitem [{\citenamefont {Ding}\ \emph {et~al.}(2020)\citenamefont {Ding},
  \citenamefont {Ross}, \citenamefont {Go}, \citenamefont {Baldrati},
  \citenamefont {Ren}, \citenamefont {Freimuth}, \citenamefont {Becker},
  \citenamefont {Kammerbauer}, \citenamefont {Yang}, \citenamefont {Jakob}
  \emph {et~al.}}]{ding2020harnessing}%
  \BibitemOpen
  \bibfield  {author} {\bibinfo {author} {\bibfnamefont {S.}~\bibnamefont
  {Ding}}, \bibinfo {author} {\bibfnamefont {A.}~\bibnamefont {Ross}}, \bibinfo
  {author} {\bibfnamefont {D.}~\bibnamefont {Go}}, \bibinfo {author}
  {\bibfnamefont {L.}~\bibnamefont {Baldrati}}, \bibinfo {author}
  {\bibfnamefont {Z.}~\bibnamefont {Ren}}, \bibinfo {author} {\bibfnamefont
  {F.}~\bibnamefont {Freimuth}}, \bibinfo {author} {\bibfnamefont
  {S.}~\bibnamefont {Becker}}, \bibinfo {author} {\bibfnamefont
  {F.}~\bibnamefont {Kammerbauer}}, \bibinfo {author} {\bibfnamefont
  {J.}~\bibnamefont {Yang}}, \bibinfo {author} {\bibfnamefont {G.}~\bibnamefont
  {Jakob}}, \emph {et~al.},\ }\bibfield  {title} {\bibinfo {title} {{Harnessing
  orbital-to-spin conversion of interfacial orbital currents for efficient
  spin-orbit torques}},\ }\href@noop {} {\bibfield  {journal} {\bibinfo
  {journal} {Physical review letters}\ }\textbf {\bibinfo {volume} {125}},\
  \bibinfo {pages} {177201} (\bibinfo {year} {2020})}\BibitemShut {NoStop}%
\bibitem [{\citenamefont {Go}\ \emph {et~al.}(2020)\citenamefont {Go},
  \citenamefont {Freimuth}, \citenamefont {Hanke}, \citenamefont {Xue},
  \citenamefont {Gomonay}, \citenamefont {Lee}, \citenamefont {Bl\"ugel},
  \citenamefont {Haney}, \citenamefont {Lee},\ and\ \citenamefont
  {Mokrousov}}]{Go2020theory}%
  \BibitemOpen
  \bibfield  {author} {\bibinfo {author} {\bibfnamefont {D.}~\bibnamefont
  {Go}}, \bibinfo {author} {\bibfnamefont {F.}~\bibnamefont {Freimuth}},
  \bibinfo {author} {\bibfnamefont {J.-P.}\ \bibnamefont {Hanke}}, \bibinfo
  {author} {\bibfnamefont {F.}~\bibnamefont {Xue}}, \bibinfo {author}
  {\bibfnamefont {O.}~\bibnamefont {Gomonay}}, \bibinfo {author} {\bibfnamefont
  {K.-J.}\ \bibnamefont {Lee}}, \bibinfo {author} {\bibfnamefont
  {S.}~\bibnamefont {Bl\"ugel}}, \bibinfo {author} {\bibfnamefont {P.~M.}\
  \bibnamefont {Haney}}, \bibinfo {author} {\bibfnamefont {H.-W.}\ \bibnamefont
  {Lee}},\ and\ \bibinfo {author} {\bibfnamefont {Y.}~\bibnamefont
  {Mokrousov}},\ }\bibfield  {title} {\bibinfo {title} {{Theory of
  current-induced angular momentum transfer dynamics in spin-orbit coupled
  systems}},\ }\href {https://doi.org/10.1103/PhysRevResearch.2.033401}
  {\bibfield  {journal} {\bibinfo  {journal} {Phys. Rev. Res.}\ }\textbf
  {\bibinfo {volume} {2}},\ \bibinfo {pages} {033401} (\bibinfo {year}
  {2020})}\BibitemShut {NoStop}%
\bibitem [{\citenamefont {Ding}\ \emph {et~al.}(2022)\citenamefont {Ding},
  \citenamefont {Liang}, \citenamefont {Go}, \citenamefont {Yun}, \citenamefont
  {Xue}, \citenamefont {Liu}, \citenamefont {Becker}, \citenamefont {Yang},
  \citenamefont {Du}, \citenamefont {Wang} \emph
  {et~al.}}]{ding2022observation}%
  \BibitemOpen
  \bibfield  {author} {\bibinfo {author} {\bibfnamefont {S.}~\bibnamefont
  {Ding}}, \bibinfo {author} {\bibfnamefont {Z.}~\bibnamefont {Liang}},
  \bibinfo {author} {\bibfnamefont {D.}~\bibnamefont {Go}}, \bibinfo {author}
  {\bibfnamefont {C.}~\bibnamefont {Yun}}, \bibinfo {author} {\bibfnamefont
  {M.}~\bibnamefont {Xue}}, \bibinfo {author} {\bibfnamefont {Z.}~\bibnamefont
  {Liu}}, \bibinfo {author} {\bibfnamefont {S.}~\bibnamefont {Becker}},
  \bibinfo {author} {\bibfnamefont {W.}~\bibnamefont {Yang}}, \bibinfo {author}
  {\bibfnamefont {H.}~\bibnamefont {Du}}, \bibinfo {author} {\bibfnamefont
  {C.}~\bibnamefont {Wang}}, \emph {et~al.},\ }\bibfield  {title} {\bibinfo
  {title} {{Observation of the orbital Rashba-Edelstein magnetoresistance}},\
  }\href@noop {} {\bibfield  {journal} {\bibinfo  {journal} {Physical review
  letters}\ }\textbf {\bibinfo {volume} {128}},\ \bibinfo {pages} {067201}
  (\bibinfo {year} {2022})}\BibitemShut {NoStop}%
\bibitem [{\citenamefont {Krishnia}\ \emph {et~al.}(2024)\citenamefont
  {Krishnia}, \citenamefont {Bony}, \citenamefont {Rongione}, \citenamefont
  {Vicente-Arche}, \citenamefont {Denneulin}, \citenamefont {Pezo},
  \citenamefont {Lu}, \citenamefont {Dunin-Borkowski}, \citenamefont {Collin},
  \citenamefont {Fert} \emph {et~al.}}]{krishnia2024quantifying}%
  \BibitemOpen
  \bibfield  {author} {\bibinfo {author} {\bibfnamefont {S.}~\bibnamefont
  {Krishnia}}, \bibinfo {author} {\bibfnamefont {B.}~\bibnamefont {Bony}},
  \bibinfo {author} {\bibfnamefont {E.}~\bibnamefont {Rongione}}, \bibinfo
  {author} {\bibfnamefont {L.~M.}\ \bibnamefont {Vicente-Arche}}, \bibinfo
  {author} {\bibfnamefont {T.}~\bibnamefont {Denneulin}}, \bibinfo {author}
  {\bibfnamefont {A.}~\bibnamefont {Pezo}}, \bibinfo {author} {\bibfnamefont
  {Y.}~\bibnamefont {Lu}}, \bibinfo {author} {\bibfnamefont {R.}~\bibnamefont
  {Dunin-Borkowski}}, \bibinfo {author} {\bibfnamefont {S.}~\bibnamefont
  {Collin}}, \bibinfo {author} {\bibfnamefont {A.}~\bibnamefont {Fert}}, \emph
  {et~al.},\ }\bibfield  {title} {\bibinfo {title} {{Quantifying the large
  contribution from orbital Rashba--Edelstein effect to the effective
  damping-like torque on magnetization}},\ }\href@noop {} {\bibfield  {journal}
  {\bibinfo  {journal} {APL Materials}\ }\textbf {\bibinfo {volume} {12}}
  (\bibinfo {year} {2024})}\BibitemShut {NoStop}%
\bibitem [{\citenamefont {Manchon}\ \emph {et~al.}(2019)\citenamefont
  {Manchon}, \citenamefont {{\v{Z}}elezn{\`y}}, \citenamefont {Miron},
  \citenamefont {Jungwirth}, \citenamefont {Sinova}, \citenamefont {Thiaville},
  \citenamefont {Garello},\ and\ \citenamefont
  {Gambardella}}]{manchon2019current}%
  \BibitemOpen
  \bibfield  {author} {\bibinfo {author} {\bibfnamefont {A.}~\bibnamefont
  {Manchon}}, \bibinfo {author} {\bibfnamefont {J.}~\bibnamefont
  {{\v{Z}}elezn{\`y}}}, \bibinfo {author} {\bibfnamefont {I.~M.}\ \bibnamefont
  {Miron}}, \bibinfo {author} {\bibfnamefont {T.}~\bibnamefont {Jungwirth}},
  \bibinfo {author} {\bibfnamefont {J.}~\bibnamefont {Sinova}}, \bibinfo
  {author} {\bibfnamefont {A.}~\bibnamefont {Thiaville}}, \bibinfo {author}
  {\bibfnamefont {K.}~\bibnamefont {Garello}},\ and\ \bibinfo {author}
  {\bibfnamefont {P.}~\bibnamefont {Gambardella}},\ }\bibfield  {title}
  {\bibinfo {title} {{Current-induced spin-orbit torques in ferromagnetic and
  antiferromagnetic systems}},\ }\href@noop {} {\bibfield  {journal} {\bibinfo
  {journal} {Reviews of Modern Physics}\ }\textbf {\bibinfo {volume} {91}},\
  \bibinfo {pages} {035004} (\bibinfo {year} {2019})}\BibitemShut {NoStop}%
\bibitem [{\citenamefont {Gupta}\ \emph {et~al.}(2024)\citenamefont {Gupta},
  \citenamefont {Chowdhury}, \citenamefont {Xu}, \citenamefont {Muduli},
  \citenamefont {Kumar}, \citenamefont {Kondou}, \citenamefont {Otani},\ and\
  \citenamefont {Muduli}}]{gupta2024generation}%
  \BibitemOpen
  \bibfield  {author} {\bibinfo {author} {\bibfnamefont {P.}~\bibnamefont
  {Gupta}}, \bibinfo {author} {\bibfnamefont {N.}~\bibnamefont {Chowdhury}},
  \bibinfo {author} {\bibfnamefont {M.}~\bibnamefont {Xu}}, \bibinfo {author}
  {\bibfnamefont {P.~K.}\ \bibnamefont {Muduli}}, \bibinfo {author}
  {\bibfnamefont {A.}~\bibnamefont {Kumar}}, \bibinfo {author} {\bibfnamefont
  {K.}~\bibnamefont {Kondou}}, \bibinfo {author} {\bibfnamefont
  {Y.}~\bibnamefont {Otani}},\ and\ \bibinfo {author} {\bibfnamefont {P.~K.}\
  \bibnamefont {Muduli}},\ }\bibfield  {title} {\bibinfo {title} {{Generation
  of out-of-plane polarized spin current in (permalloy, Cu)/EuS interfaces}},\
  }\href@noop {} {\bibfield  {journal} {\bibinfo  {journal} {Physical Review
  B}\ }\textbf {\bibinfo {volume} {109}},\ \bibinfo {pages} {L060405} (\bibinfo
  {year} {2024})}\BibitemShut {NoStop}%
\bibitem [{\citenamefont {Freimuth}\ \emph {et~al.}(2015)\citenamefont
  {Freimuth}, \citenamefont {Bl{\"u}gel},\ and\ \citenamefont
  {Mokrousov}}]{freimuth2015direct}%
  \BibitemOpen
  \bibfield  {author} {\bibinfo {author} {\bibfnamefont {F.}~\bibnamefont
  {Freimuth}}, \bibinfo {author} {\bibfnamefont {S.}~\bibnamefont
  {Bl{\"u}gel}},\ and\ \bibinfo {author} {\bibfnamefont {Y.}~\bibnamefont
  {Mokrousov}},\ }\bibfield  {title} {\bibinfo {title} {{Direct and inverse
  spin-orbit torques}},\ }\href@noop {} {\bibfield  {journal} {\bibinfo
  {journal} {Physical Review B}\ }\textbf {\bibinfo {volume} {92}},\ \bibinfo
  {pages} {064415} (\bibinfo {year} {2015})}\BibitemShut {NoStop}%
\bibitem [{\citenamefont {Go}\ \emph {et~al.}(2023{\natexlab{b}})\citenamefont
  {Go}, \citenamefont {Ando}, \citenamefont {Pezo}, \citenamefont {Bl{\"u}gel},
  \citenamefont {Manchon},\ and\ \citenamefont {Mokrousov}}]{go2023orbital}%
  \BibitemOpen
  \bibfield  {author} {\bibinfo {author} {\bibfnamefont {D.}~\bibnamefont
  {Go}}, \bibinfo {author} {\bibfnamefont {K.}~\bibnamefont {Ando}}, \bibinfo
  {author} {\bibfnamefont {A.}~\bibnamefont {Pezo}}, \bibinfo {author}
  {\bibfnamefont {S.}~\bibnamefont {Bl{\"u}gel}}, \bibinfo {author}
  {\bibfnamefont {A.}~\bibnamefont {Manchon}},\ and\ \bibinfo {author}
  {\bibfnamefont {Y.}~\bibnamefont {Mokrousov}},\ }\bibfield  {title} {\bibinfo
  {title} {{Orbital Pumping by Magnetization Dynamics in Ferromagnets}},\
  }\href@noop {} {\bibfield  {journal} {\bibinfo  {journal} {arXiv preprint
  arXiv:2309.14817}\ } (\bibinfo {year} {2023}{\natexlab{b}})}\BibitemShut
  {NoStop}%
\bibitem [{\citenamefont {Han}\ \emph {et~al.}(2025)\citenamefont {Han},
  \citenamefont {Ko}, \citenamefont {Oh}, \citenamefont {Lee}, \citenamefont
  {Lee},\ and\ \citenamefont {Kim}}]{han2024orbital}%
  \BibitemOpen
  \bibfield  {author} {\bibinfo {author} {\bibfnamefont {S.}~\bibnamefont
  {Han}}, \bibinfo {author} {\bibfnamefont {H.-W.}\ \bibnamefont {Ko}},
  \bibinfo {author} {\bibfnamefont {J.~H.}\ \bibnamefont {Oh}}, \bibinfo
  {author} {\bibfnamefont {H.-W.}\ \bibnamefont {Lee}}, \bibinfo {author}
  {\bibfnamefont {K.-J.}\ \bibnamefont {Lee}},\ and\ \bibinfo {author}
  {\bibfnamefont {K.-W.}\ \bibnamefont {Kim}},\ }\bibfield  {title} {\bibinfo
  {title} {Orbital pumping incorporating both orbital angular momentum and
  position},\ }\href {https://doi.org/10.1103/PhysRevLett.134.036305}
  {\bibfield  {journal} {\bibinfo  {journal} {Phys. Rev. Lett.}\ }\textbf
  {\bibinfo {volume} {134}},\ \bibinfo {pages} {036305} (\bibinfo {year}
  {2025})}\BibitemShut {NoStop}%
\bibitem [{\citenamefont {Hayashi}\ \emph {et~al.}(2024)\citenamefont
  {Hayashi}, \citenamefont {Go}, \citenamefont {Haku}, \citenamefont
  {Mokrousov},\ and\ \citenamefont {Ando}}]{hayashi2024observation}%
  \BibitemOpen
  \bibfield  {author} {\bibinfo {author} {\bibfnamefont {H.}~\bibnamefont
  {Hayashi}}, \bibinfo {author} {\bibfnamefont {D.}~\bibnamefont {Go}},
  \bibinfo {author} {\bibfnamefont {S.}~\bibnamefont {Haku}}, \bibinfo {author}
  {\bibfnamefont {Y.}~\bibnamefont {Mokrousov}},\ and\ \bibinfo {author}
  {\bibfnamefont {K.}~\bibnamefont {Ando}},\ }\bibfield  {title} {\bibinfo
  {title} {Observation of orbital pumping},\ }\href@noop {} {\bibfield
  {journal} {\bibinfo  {journal} {Nature Electronics}\ }\textbf {\bibinfo
  {volume} {7}},\ \bibinfo {pages} {646} (\bibinfo {year} {2024})}\BibitemShut
  {NoStop}%
\bibitem [{\citenamefont {Ding}\ \emph {et~al.}(2024)\citenamefont {Ding},
  \citenamefont {Kang}, \citenamefont {Legrand},\ and\ \citenamefont
  {Gambardella}}]{ding2024orbital}%
  \BibitemOpen
  \bibfield  {author} {\bibinfo {author} {\bibfnamefont {S.}~\bibnamefont
  {Ding}}, \bibinfo {author} {\bibfnamefont {M.-G.}\ \bibnamefont {Kang}},
  \bibinfo {author} {\bibfnamefont {W.}~\bibnamefont {Legrand}},\ and\ \bibinfo
  {author} {\bibfnamefont {P.}~\bibnamefont {Gambardella}},\ }\bibfield
  {title} {\bibinfo {title} {{Orbital Torque in Rare-Earth Transition-Metal
  Ferrimagnets}},\ }\href@noop {} {\bibfield  {journal} {\bibinfo  {journal}
  {Physical Review Letters}\ }\textbf {\bibinfo {volume} {132}},\ \bibinfo
  {pages} {236702} (\bibinfo {year} {2024})}\BibitemShut {NoStop}%
\bibitem [{\citenamefont {Liu}\ \emph {et~al.}(2024)\citenamefont {Liu},
  \citenamefont {Jiang}, \citenamefont {Zhao}, \citenamefont {Chen},
  \citenamefont {Lai}, \citenamefont {Liu},\ and\ \citenamefont
  {Zhang}}]{liu2024qualitative}%
  \BibitemOpen
  \bibfield  {author} {\bibinfo {author} {\bibfnamefont {L.}~\bibnamefont
  {Liu}}, \bibinfo {author} {\bibfnamefont {T.}~\bibnamefont {Jiang}}, \bibinfo
  {author} {\bibfnamefont {X.}~\bibnamefont {Zhao}}, \bibinfo {author}
  {\bibfnamefont {K.}~\bibnamefont {Chen}}, \bibinfo {author} {\bibfnamefont
  {T.}~\bibnamefont {Lai}}, \bibinfo {author} {\bibfnamefont {W.}~\bibnamefont
  {Liu}},\ and\ \bibinfo {author} {\bibfnamefont {Z.}~\bibnamefont {Zhang}},\
  }\bibfield  {title} {\bibinfo {title} {{Qualitative Identification of the
  Spin-to-Orbital Conversion Mechanism Modulated by Rare-Earth Nd, Gd, and Ho
  Metals via Terahertz Emission Measurements}},\ }\href@noop {} {\bibfield
  {journal} {\bibinfo  {journal} {Advanced Functional Materials}\ ,\ \bibinfo
  {pages} {2411262}} (\bibinfo {year} {2024})}\BibitemShut {NoStop}%
\bibitem [{\citenamefont {Gibertini}\ \emph {et~al.}(2019)\citenamefont
  {Gibertini}, \citenamefont {Koperski}, \citenamefont {Morpurgo},\ and\
  \citenamefont {Novoselov}}]{gibertini2019magnetic}%
  \BibitemOpen
  \bibfield  {author} {\bibinfo {author} {\bibfnamefont {M.}~\bibnamefont
  {Gibertini}}, \bibinfo {author} {\bibfnamefont {M.}~\bibnamefont {Koperski}},
  \bibinfo {author} {\bibfnamefont {A.~F.}\ \bibnamefont {Morpurgo}},\ and\
  \bibinfo {author} {\bibfnamefont {K.~S.}\ \bibnamefont {Novoselov}},\
  }\bibfield  {title} {\bibinfo {title} {{Magnetic 2D materials and
  heterostructures}},\ }\href@noop {} {\bibfield  {journal} {\bibinfo
  {journal} {Nature nanotechnology}\ }\textbf {\bibinfo {volume} {14}},\
  \bibinfo {pages} {408} (\bibinfo {year} {2019})}\BibitemShut {NoStop}%
\bibitem [{\citenamefont {Yi}\ \emph {et~al.}(2019)\citenamefont {Yi},
  \citenamefont {Chen}, \citenamefont {Yu}, \citenamefont {Zhou},\ and\
  \citenamefont {Li}}]{yi2019recent}%
  \BibitemOpen
  \bibfield  {author} {\bibinfo {author} {\bibfnamefont {Y.}~\bibnamefont
  {Yi}}, \bibinfo {author} {\bibfnamefont {Z.}~\bibnamefont {Chen}}, \bibinfo
  {author} {\bibfnamefont {X.-F.}\ \bibnamefont {Yu}}, \bibinfo {author}
  {\bibfnamefont {Z.-K.}\ \bibnamefont {Zhou}},\ and\ \bibinfo {author}
  {\bibfnamefont {J.}~\bibnamefont {Li}},\ }\bibfield  {title} {\bibinfo
  {title} {{Recent advances in quantum effects of 2D materials}},\ }\href@noop
  {} {\bibfield  {journal} {\bibinfo  {journal} {Advanced Quantum
  Technologies}\ }\textbf {\bibinfo {volume} {2}},\ \bibinfo {pages} {1800111}
  (\bibinfo {year} {2019})}\BibitemShut {NoStop}%
\bibitem [{\citenamefont {Ahn}(2020)}]{ahn20202d}%
  \BibitemOpen
  \bibfield  {author} {\bibinfo {author} {\bibfnamefont {E.~C.}\ \bibnamefont
  {Ahn}},\ }\bibfield  {title} {\bibinfo {title} {{2D materials for spintronic
  devices}},\ }\href@noop {} {\bibfield  {journal} {\bibinfo  {journal} {npj 2D
  Materials and Applications}\ }\textbf {\bibinfo {volume} {4}},\ \bibinfo
  {pages} {17} (\bibinfo {year} {2020})}\BibitemShut {NoStop}%
\bibitem [{\citenamefont {Husain}\ \emph
  {et~al.}(2020{\natexlab{a}})\citenamefont {Husain}, \citenamefont {Gupta},
  \citenamefont {Kumar}, \citenamefont {Kumar}, \citenamefont {Behera},
  \citenamefont {Brucas}, \citenamefont {Chaudhary},\ and\ \citenamefont
  {Svedlindh}}]{husain2020emergence}%
  \BibitemOpen
  \bibfield  {author} {\bibinfo {author} {\bibfnamefont {S.}~\bibnamefont
  {Husain}}, \bibinfo {author} {\bibfnamefont {R.}~\bibnamefont {Gupta}},
  \bibinfo {author} {\bibfnamefont {A.}~\bibnamefont {Kumar}}, \bibinfo
  {author} {\bibfnamefont {P.}~\bibnamefont {Kumar}}, \bibinfo {author}
  {\bibfnamefont {N.}~\bibnamefont {Behera}}, \bibinfo {author} {\bibfnamefont
  {R.}~\bibnamefont {Brucas}}, \bibinfo {author} {\bibfnamefont
  {S.}~\bibnamefont {Chaudhary}},\ and\ \bibinfo {author} {\bibfnamefont
  {P.}~\bibnamefont {Svedlindh}},\ }\bibfield  {title} {\bibinfo {title}
  {{Emergence of spin--orbit torques in 2D transition metal dichalcogenides: A
  status update}},\ }\href@noop {} {\bibfield  {journal} {\bibinfo  {journal}
  {Applied Physics Reviews}\ }\textbf {\bibinfo {volume} {7}} (\bibinfo {year}
  {2020}{\natexlab{a}})}\BibitemShut {NoStop}%
\bibitem [{\citenamefont {Grytsiuk}\ \emph {et~al.}(2020)\citenamefont
  {Grytsiuk}, \citenamefont {Hanke}, \citenamefont {Hoffmann}, \citenamefont
  {Bouaziz}, \citenamefont {Gomonay}, \citenamefont {Bihlmayer}, \citenamefont
  {Lounis}, \citenamefont {Mokrousov},\ and\ \citenamefont
  {Bl{\"u}gel}}]{grytsiuk2020topological}%
  \BibitemOpen
  \bibfield  {author} {\bibinfo {author} {\bibfnamefont {S.}~\bibnamefont
  {Grytsiuk}}, \bibinfo {author} {\bibfnamefont {J.-P.}\ \bibnamefont {Hanke}},
  \bibinfo {author} {\bibfnamefont {M.}~\bibnamefont {Hoffmann}}, \bibinfo
  {author} {\bibfnamefont {J.}~\bibnamefont {Bouaziz}}, \bibinfo {author}
  {\bibfnamefont {O.}~\bibnamefont {Gomonay}}, \bibinfo {author} {\bibfnamefont
  {G.}~\bibnamefont {Bihlmayer}}, \bibinfo {author} {\bibfnamefont
  {S.}~\bibnamefont {Lounis}}, \bibinfo {author} {\bibfnamefont
  {Y.}~\bibnamefont {Mokrousov}},\ and\ \bibinfo {author} {\bibfnamefont
  {S.}~\bibnamefont {Bl{\"u}gel}},\ }\bibfield  {title} {\bibinfo {title}
  {{Topological--chiral magnetic interactions driven by emergent orbital
  magnetism}},\ }\href@noop {} {\bibfield  {journal} {\bibinfo  {journal}
  {Nature communications}\ }\textbf {\bibinfo {volume} {11}},\ \bibinfo {pages}
  {511} (\bibinfo {year} {2020})}\BibitemShut {NoStop}%
\bibitem [{\citenamefont {Niu}\ \emph {et~al.}(2019)\citenamefont {Niu},
  \citenamefont {Hanke}, \citenamefont {Buhl}, \citenamefont {Zhang},
  \citenamefont {Plucinski}, \citenamefont {Wortmann}, \citenamefont
  {Bl{\"u}gel}, \citenamefont {Bihlmayer},\ and\ \citenamefont
  {Mokrousov}}]{niu2019mixed}%
  \BibitemOpen
  \bibfield  {author} {\bibinfo {author} {\bibfnamefont {C.}~\bibnamefont
  {Niu}}, \bibinfo {author} {\bibfnamefont {J.-P.}\ \bibnamefont {Hanke}},
  \bibinfo {author} {\bibfnamefont {P.~M.}\ \bibnamefont {Buhl}}, \bibinfo
  {author} {\bibfnamefont {H.}~\bibnamefont {Zhang}}, \bibinfo {author}
  {\bibfnamefont {L.}~\bibnamefont {Plucinski}}, \bibinfo {author}
  {\bibfnamefont {D.}~\bibnamefont {Wortmann}}, \bibinfo {author}
  {\bibfnamefont {S.}~\bibnamefont {Bl{\"u}gel}}, \bibinfo {author}
  {\bibfnamefont {G.}~\bibnamefont {Bihlmayer}},\ and\ \bibinfo {author}
  {\bibfnamefont {Y.}~\bibnamefont {Mokrousov}},\ }\bibfield  {title} {\bibinfo
  {title} {{Mixed topological semimetals driven by orbital complexity in
  two-dimensional ferromagnets}},\ }\href@noop {} {\bibfield  {journal}
  {\bibinfo  {journal} {Nature communications}\ }\textbf {\bibinfo {volume}
  {10}},\ \bibinfo {pages} {3179} (\bibinfo {year} {2019})}\BibitemShut
  {NoStop}%
\bibitem [{\citenamefont {Cysne}\ \emph {et~al.}(2021)\citenamefont {Cysne},
  \citenamefont {Costa}, \citenamefont {Canonico}, \citenamefont {Nardelli},
  \citenamefont {Muniz},\ and\ \citenamefont
  {Rappoport}}]{cysne2021disentangling}%
  \BibitemOpen
  \bibfield  {author} {\bibinfo {author} {\bibfnamefont {T.~P.}\ \bibnamefont
  {Cysne}}, \bibinfo {author} {\bibfnamefont {M.}~\bibnamefont {Costa}},
  \bibinfo {author} {\bibfnamefont {L.~M.}\ \bibnamefont {Canonico}}, \bibinfo
  {author} {\bibfnamefont {M.~B.}\ \bibnamefont {Nardelli}}, \bibinfo {author}
  {\bibfnamefont {R.}~\bibnamefont {Muniz}},\ and\ \bibinfo {author}
  {\bibfnamefont {T.~G.}\ \bibnamefont {Rappoport}},\ }\bibfield  {title}
  {\bibinfo {title} {{Disentangling orbital and valley Hall effects in bilayers
  of transition metal dichalcogenides}},\ }\href@noop {} {\bibfield  {journal}
  {\bibinfo  {journal} {Physical review letters}\ }\textbf {\bibinfo {volume}
  {126}},\ \bibinfo {pages} {056601} (\bibinfo {year} {2021})}\BibitemShut
  {NoStop}%
\bibitem [{\citenamefont {Canonico}\ \emph {et~al.}(2020)\citenamefont
  {Canonico}, \citenamefont {Cysne}, \citenamefont {Molina-Sanchez},
  \citenamefont {Muniz},\ and\ \citenamefont
  {Rappoport}}]{canonico2020orbital}%
  \BibitemOpen
  \bibfield  {author} {\bibinfo {author} {\bibfnamefont {L.~M.}\ \bibnamefont
  {Canonico}}, \bibinfo {author} {\bibfnamefont {T.~P.}\ \bibnamefont {Cysne}},
  \bibinfo {author} {\bibfnamefont {A.}~\bibnamefont {Molina-Sanchez}},
  \bibinfo {author} {\bibfnamefont {R.}~\bibnamefont {Muniz}},\ and\ \bibinfo
  {author} {\bibfnamefont {T.~G.}\ \bibnamefont {Rappoport}},\ }\bibfield
  {title} {\bibinfo {title} {{Orbital Hall insulating phase in transition metal
  dichalcogenide monolayers}},\ }\href@noop {} {\bibfield  {journal} {\bibinfo
  {journal} {Physical Review B}\ }\textbf {\bibinfo {volume} {101}},\ \bibinfo
  {pages} {161409} (\bibinfo {year} {2020})}\BibitemShut {NoStop}%
\bibitem [{\citenamefont {Fan}\ \emph {et~al.}(2024)\citenamefont {Fan},
  \citenamefont {Xiao},\ and\ \citenamefont {Yao}}]{fan2024orbital}%
  \BibitemOpen
  \bibfield  {author} {\bibinfo {author} {\bibfnamefont {F.-R.}\ \bibnamefont
  {Fan}}, \bibinfo {author} {\bibfnamefont {C.}~\bibnamefont {Xiao}},\ and\
  \bibinfo {author} {\bibfnamefont {W.}~\bibnamefont {Yao}},\ }\bibfield
  {title} {\bibinfo {title} {{Orbital Chern insulator at $\nu$=- 2 in twisted
  MoTe 2}},\ }\href@noop {} {\bibfield  {journal} {\bibinfo  {journal}
  {Physical Review B}\ }\textbf {\bibinfo {volume} {109}},\ \bibinfo {pages}
  {L041403} (\bibinfo {year} {2024})}\BibitemShut {NoStop}%
\bibitem [{\citenamefont {Xue}\ \emph {et~al.}(2020)\citenamefont {Xue},
  \citenamefont {Amin},\ and\ \citenamefont {Haney}}]{xue2020imaging}%
  \BibitemOpen
  \bibfield  {author} {\bibinfo {author} {\bibfnamefont {F.}~\bibnamefont
  {Xue}}, \bibinfo {author} {\bibfnamefont {V.}~\bibnamefont {Amin}},\ and\
  \bibinfo {author} {\bibfnamefont {P.~M.}\ \bibnamefont {Haney}},\ }\bibfield
  {title} {\bibinfo {title} {{Imaging the valley and orbital Hall effect in
  monolayer MoS 2}},\ }\href@noop {} {\bibfield  {journal} {\bibinfo  {journal}
  {Physical Review B}\ }\textbf {\bibinfo {volume} {102}},\ \bibinfo {pages}
  {161103} (\bibinfo {year} {2020})}\BibitemShut {NoStop}%
\bibitem [{\citenamefont {Lee}\ \emph {et~al.}(2016)\citenamefont {Lee},
  \citenamefont {Mak},\ and\ \citenamefont {Shan}}]{lee2016electrical}%
  \BibitemOpen
  \bibfield  {author} {\bibinfo {author} {\bibfnamefont {J.}~\bibnamefont
  {Lee}}, \bibinfo {author} {\bibfnamefont {K.~F.}\ \bibnamefont {Mak}},\ and\
  \bibinfo {author} {\bibfnamefont {J.}~\bibnamefont {Shan}},\ }\bibfield
  {title} {\bibinfo {title} {{Electrical control of the valley Hall effect in
  bilayer MoS2 transistors}},\ }\href@noop {} {\bibfield  {journal} {\bibinfo
  {journal} {Nature nanotechnology}\ }\textbf {\bibinfo {volume} {11}},\
  \bibinfo {pages} {421} (\bibinfo {year} {2016})}\BibitemShut {NoStop}%
\bibitem [{\citenamefont {Mak}\ \emph {et~al.}(2014)\citenamefont {Mak},
  \citenamefont {McGill}, \citenamefont {Park},\ and\ \citenamefont
  {McEuen}}]{mak2014valley}%
  \BibitemOpen
  \bibfield  {author} {\bibinfo {author} {\bibfnamefont {K.~F.}\ \bibnamefont
  {Mak}}, \bibinfo {author} {\bibfnamefont {K.~L.}\ \bibnamefont {McGill}},
  \bibinfo {author} {\bibfnamefont {J.}~\bibnamefont {Park}},\ and\ \bibinfo
  {author} {\bibfnamefont {P.~L.}\ \bibnamefont {McEuen}},\ }\bibfield  {title}
  {\bibinfo {title} {{The valley Hall effect in MoS2 transistors}},\
  }\href@noop {} {\bibfield  {journal} {\bibinfo  {journal} {Science}\ }\textbf
  {\bibinfo {volume} {344}},\ \bibinfo {pages} {1489} (\bibinfo {year}
  {2014})}\BibitemShut {NoStop}%
\bibitem [{\citenamefont {Bhowal}\ and\ \citenamefont
  {Vignale}(2021)}]{bhowal2021orbital}%
  \BibitemOpen
  \bibfield  {author} {\bibinfo {author} {\bibfnamefont {S.}~\bibnamefont
  {Bhowal}}\ and\ \bibinfo {author} {\bibfnamefont {G.}~\bibnamefont
  {Vignale}},\ }\bibfield  {title} {\bibinfo {title} {Orbital hall effect as an
  alternative to valley hall effect in gapped graphene},\ }\href
  {https://doi.org/10.1103/PhysRevB.103.195309} {\bibfield  {journal} {\bibinfo
   {journal} {Phys. Rev. B}\ }\textbf {\bibinfo {volume} {103}},\ \bibinfo
  {pages} {195309} (\bibinfo {year} {2021})}\BibitemShut {NoStop}%
\bibitem [{\citenamefont {Zeer}\ \emph {et~al.}(2022)\citenamefont {Zeer},
  \citenamefont {Go}, \citenamefont {Carbone}, \citenamefont {Saunderson},
  \citenamefont {Redies}, \citenamefont {Kl{\"a}ui}, \citenamefont {Ghabboun},
  \citenamefont {Wulfhekel}, \citenamefont {Bl{\"u}gel},\ and\ \citenamefont
  {Mokrousov}}]{zeer2022spin}%
  \BibitemOpen
  \bibfield  {author} {\bibinfo {author} {\bibfnamefont {M.}~\bibnamefont
  {Zeer}}, \bibinfo {author} {\bibfnamefont {D.}~\bibnamefont {Go}}, \bibinfo
  {author} {\bibfnamefont {J.~P.}\ \bibnamefont {Carbone}}, \bibinfo {author}
  {\bibfnamefont {T.~G.}\ \bibnamefont {Saunderson}}, \bibinfo {author}
  {\bibfnamefont {M.}~\bibnamefont {Redies}}, \bibinfo {author} {\bibfnamefont
  {M.}~\bibnamefont {Kl{\"a}ui}}, \bibinfo {author} {\bibfnamefont
  {J.}~\bibnamefont {Ghabboun}}, \bibinfo {author} {\bibfnamefont
  {W.}~\bibnamefont {Wulfhekel}}, \bibinfo {author} {\bibfnamefont
  {S.}~\bibnamefont {Bl{\"u}gel}},\ and\ \bibinfo {author} {\bibfnamefont
  {Y.}~\bibnamefont {Mokrousov}},\ }\bibfield  {title} {\bibinfo {title} {{Spin
  and orbital transport in rare-earth dichalcogenides: The case of EuS 2}},\
  }\href@noop {} {\bibfield  {journal} {\bibinfo  {journal} {Physical review
  materials}\ }\textbf {\bibinfo {volume} {6}},\ \bibinfo {pages} {074004}
  (\bibinfo {year} {2022})}\BibitemShut {NoStop}%
\bibitem [{\citenamefont {Zeer}\ \emph {et~al.}(2024)\citenamefont {Zeer},
  \citenamefont {Go}, \citenamefont {Schmitz}, \citenamefont {Saunderson},
  \citenamefont {Wang}, \citenamefont {Ghabboun}, \citenamefont {Bl{\"u}gel},
  \citenamefont {Wulfhekel},\ and\ \citenamefont
  {Mokrousov}}]{zeer2024promoting}%
  \BibitemOpen
  \bibfield  {author} {\bibinfo {author} {\bibfnamefont {M.}~\bibnamefont
  {Zeer}}, \bibinfo {author} {\bibfnamefont {D.}~\bibnamefont {Go}}, \bibinfo
  {author} {\bibfnamefont {P.}~\bibnamefont {Schmitz}}, \bibinfo {author}
  {\bibfnamefont {T.~G.}\ \bibnamefont {Saunderson}}, \bibinfo {author}
  {\bibfnamefont {H.}~\bibnamefont {Wang}}, \bibinfo {author} {\bibfnamefont
  {J.}~\bibnamefont {Ghabboun}}, \bibinfo {author} {\bibfnamefont
  {S.}~\bibnamefont {Bl{\"u}gel}}, \bibinfo {author} {\bibfnamefont
  {W.}~\bibnamefont {Wulfhekel}},\ and\ \bibinfo {author} {\bibfnamefont
  {Y.}~\bibnamefont {Mokrousov}},\ }\bibfield  {title} {\bibinfo {title}
  {{Promoting p-based Hall effects by p-d-f hybridization in Gd-based
  dichalcogenides}},\ }\href@noop {} {\bibfield  {journal} {\bibinfo  {journal}
  {Physical Review Research}\ }\textbf {\bibinfo {volume} {6}},\ \bibinfo
  {pages} {013095} (\bibinfo {year} {2024})}\BibitemShut {NoStop}%
\bibitem [{\citenamefont {Hidding}\ and\ \citenamefont
  {Guimar{\~a}es}(2020)}]{hidding2020spin}%
  \BibitemOpen
  \bibfield  {author} {\bibinfo {author} {\bibfnamefont {J.}~\bibnamefont
  {Hidding}}\ and\ \bibinfo {author} {\bibfnamefont {M.~H.}\ \bibnamefont
  {Guimar{\~a}es}},\ }\bibfield  {title} {\bibinfo {title} {Spin-orbit torques
  in transition metal dichalcogenide/ferromagnet heterostructures},\
  }\href@noop {} {\bibfield  {journal} {\bibinfo  {journal} {Frontiers in
  Materials}\ }\textbf {\bibinfo {volume} {7}},\ \bibinfo {pages} {594771}
  (\bibinfo {year} {2020})}\BibitemShut {NoStop}%
\bibitem [{\citenamefont {Manchon}\ and\ \citenamefont
  {Zhang}(2009)}]{manchon2009theory}%
  \BibitemOpen
  \bibfield  {author} {\bibinfo {author} {\bibfnamefont {A.}~\bibnamefont
  {Manchon}}\ and\ \bibinfo {author} {\bibfnamefont {S.}~\bibnamefont
  {Zhang}},\ }\bibfield  {title} {\bibinfo {title} {{Theory of spin torque due
  to spin-orbit coupling}},\ }\href@noop {} {\bibfield  {journal} {\bibinfo
  {journal} {Physical Review B}\ }\textbf {\bibinfo {volume} {79}},\ \bibinfo
  {pages} {094422} (\bibinfo {year} {2009})}\BibitemShut {NoStop}%
\bibitem [{\citenamefont {Saunderson}\ \emph {et~al.}(2022)\citenamefont
  {Saunderson}, \citenamefont {Go}, \citenamefont {Bl{\"u}gel}, \citenamefont
  {Kl{\"a}ui},\ and\ \citenamefont {Mokrousov}}]{saunderson2022hidden}%
  \BibitemOpen
  \bibfield  {author} {\bibinfo {author} {\bibfnamefont {T.~G.}\ \bibnamefont
  {Saunderson}}, \bibinfo {author} {\bibfnamefont {D.}~\bibnamefont {Go}},
  \bibinfo {author} {\bibfnamefont {S.}~\bibnamefont {Bl{\"u}gel}}, \bibinfo
  {author} {\bibfnamefont {M.}~\bibnamefont {Kl{\"a}ui}},\ and\ \bibinfo
  {author} {\bibfnamefont {Y.}~\bibnamefont {Mokrousov}},\ }\bibfield  {title}
  {\bibinfo {title} {{Hidden interplay of current-induced spin and orbital
  torques in bulk Fe 3 GeTe 2}},\ }\href@noop {} {\bibfield  {journal}
  {\bibinfo  {journal} {Physical Review Research}\ }\textbf {\bibinfo {volume}
  {4}},\ \bibinfo {pages} {L042022} (\bibinfo {year} {2022})}\BibitemShut
  {NoStop}%
\bibitem [{\citenamefont {Guo}\ \emph {et~al.}(2018)\citenamefont {Guo},
  \citenamefont {Zhang}, \citenamefont {Zeng}, \citenamefont {Da},
  \citenamefont {Yan}, \citenamefont {Liu},\ and\ \citenamefont
  {Mou}}]{guo2018}%
  \BibitemOpen
  \bibfield  {author} {\bibinfo {author} {\bibfnamefont {Y.-D.}\ \bibnamefont
  {Guo}}, \bibinfo {author} {\bibfnamefont {H.-B.}\ \bibnamefont {Zhang}},
  \bibinfo {author} {\bibfnamefont {H.-L.}\ \bibnamefont {Zeng}}, \bibinfo
  {author} {\bibfnamefont {H.-X.}\ \bibnamefont {Da}}, \bibinfo {author}
  {\bibfnamefont {X.-H.}\ \bibnamefont {Yan}}, \bibinfo {author} {\bibfnamefont
  {W.-Y.}\ \bibnamefont {Liu}},\ and\ \bibinfo {author} {\bibfnamefont {X.-Y.}\
  \bibnamefont {Mou}},\ }\bibfield  {title} {\bibinfo {title} {{A progressive
  metal-semiconductor transition in two-faced Janus monolayer transition-metal
  chalcogenides}},\ }\href@noop {} {\bibfield  {journal} {\bibinfo  {journal}
  {Physical Chemistry Chemical Physics}\ }\textbf {\bibinfo {volume} {20}},\
  \bibinfo {pages} {21113} (\bibinfo {year} {2018})}\BibitemShut {NoStop}%
\bibitem [{\citenamefont {Zhang}\ \emph
  {et~al.}(2020{\natexlab{a}})\citenamefont {Zhang}, \citenamefont {Yang},
  \citenamefont {Gong}, \citenamefont {Pan}, \citenamefont {Wang},
  \citenamefont {Guo}, \citenamefont {Zhang},\ and\ \citenamefont
  {Fu}}]{zhang2020recent}%
  \BibitemOpen
  \bibfield  {author} {\bibinfo {author} {\bibfnamefont {L.}~\bibnamefont
  {Zhang}}, \bibinfo {author} {\bibfnamefont {Z.}~\bibnamefont {Yang}},
  \bibinfo {author} {\bibfnamefont {T.}~\bibnamefont {Gong}}, \bibinfo {author}
  {\bibfnamefont {R.}~\bibnamefont {Pan}}, \bibinfo {author} {\bibfnamefont
  {H.}~\bibnamefont {Wang}}, \bibinfo {author} {\bibfnamefont {Z.}~\bibnamefont
  {Guo}}, \bibinfo {author} {\bibfnamefont {H.}~\bibnamefont {Zhang}},\ and\
  \bibinfo {author} {\bibfnamefont {X.}~\bibnamefont {Fu}},\ }\bibfield
  {title} {\bibinfo {title} {{Recent advances in emerging Janus two-dimensional
  materials: from fundamental physics to device applications}},\ }\href@noop {}
  {\bibfield  {journal} {\bibinfo  {journal} {Journal of materials chemistry
  A}\ }\textbf {\bibinfo {volume} {8}},\ \bibinfo {pages} {8813} (\bibinfo
  {year} {2020}{\natexlab{a}})}\BibitemShut {NoStop}%
\bibitem [{\citenamefont {Zhang}\ \emph {et~al.}(2017)\citenamefont {Zhang},
  \citenamefont {Jia}, \citenamefont {Kholmanov}, \citenamefont {Dong},
  \citenamefont {Er}, \citenamefont {Chen}, \citenamefont {Guo}, \citenamefont
  {Jin}, \citenamefont {Shenoy}, \citenamefont {Shi} \emph
  {et~al.}}]{zhang2017}%
  \BibitemOpen
  \bibfield  {author} {\bibinfo {author} {\bibfnamefont {J.}~\bibnamefont
  {Zhang}}, \bibinfo {author} {\bibfnamefont {S.}~\bibnamefont {Jia}}, \bibinfo
  {author} {\bibfnamefont {I.}~\bibnamefont {Kholmanov}}, \bibinfo {author}
  {\bibfnamefont {L.}~\bibnamefont {Dong}}, \bibinfo {author} {\bibfnamefont
  {D.}~\bibnamefont {Er}}, \bibinfo {author} {\bibfnamefont {W.}~\bibnamefont
  {Chen}}, \bibinfo {author} {\bibfnamefont {H.}~\bibnamefont {Guo}}, \bibinfo
  {author} {\bibfnamefont {Z.}~\bibnamefont {Jin}}, \bibinfo {author}
  {\bibfnamefont {V.~B.}\ \bibnamefont {Shenoy}}, \bibinfo {author}
  {\bibfnamefont {L.}~\bibnamefont {Shi}}, \emph {et~al.},\ }\bibfield  {title}
  {\bibinfo {title} {{Janus monolayer transition-metal dichalcogenides}},\
  }\href@noop {} {\bibfield  {journal} {\bibinfo  {journal} {ACS nano}\
  }\textbf {\bibinfo {volume} {11}},\ \bibinfo {pages} {8192} (\bibinfo {year}
  {2017})}\BibitemShut {NoStop}%
\bibitem [{\citenamefont {Lu}\ \emph {et~al.}(2017)\citenamefont {Lu},
  \citenamefont {Zhu}, \citenamefont {Xiao}, \citenamefont {Chuu},
  \citenamefont {Han}, \citenamefont {Chiu}, \citenamefont {Cheng},
  \citenamefont {Yang}, \citenamefont {Wei}, \citenamefont {Yang} \emph
  {et~al.}}]{lu2017}%
  \BibitemOpen
  \bibfield  {author} {\bibinfo {author} {\bibfnamefont {A.-Y.}\ \bibnamefont
  {Lu}}, \bibinfo {author} {\bibfnamefont {H.}~\bibnamefont {Zhu}}, \bibinfo
  {author} {\bibfnamefont {J.}~\bibnamefont {Xiao}}, \bibinfo {author}
  {\bibfnamefont {C.-P.}\ \bibnamefont {Chuu}}, \bibinfo {author}
  {\bibfnamefont {Y.}~\bibnamefont {Han}}, \bibinfo {author} {\bibfnamefont
  {M.-H.}\ \bibnamefont {Chiu}}, \bibinfo {author} {\bibfnamefont {C.-C.}\
  \bibnamefont {Cheng}}, \bibinfo {author} {\bibfnamefont {C.-W.}\ \bibnamefont
  {Yang}}, \bibinfo {author} {\bibfnamefont {K.-H.}\ \bibnamefont {Wei}},
  \bibinfo {author} {\bibfnamefont {Y.}~\bibnamefont {Yang}}, \emph {et~al.},\
  }\bibfield  {title} {\bibinfo {title} {{Janus monolayers of transition metal
  dichalcogenides}},\ }\href@noop {} {\bibfield  {journal} {\bibinfo  {journal}
  {Nature nanotechnology}\ }\textbf {\bibinfo {volume} {12}},\ \bibinfo {pages}
  {744} (\bibinfo {year} {2017})}\BibitemShut {NoStop}%
\bibitem [{\citenamefont {Zhang}\ \emph
  {et~al.}(2020{\natexlab{b}})\citenamefont {Zhang}, \citenamefont {Yang},
  \citenamefont {Gong}, \citenamefont {Pan}, \citenamefont {Wang},
  \citenamefont {Guo}, \citenamefont {Zhang},\ and\ \citenamefont
  {Fu}}]{zhang2020}%
  \BibitemOpen
  \bibfield  {author} {\bibinfo {author} {\bibfnamefont {L.}~\bibnamefont
  {Zhang}}, \bibinfo {author} {\bibfnamefont {Z.}~\bibnamefont {Yang}},
  \bibinfo {author} {\bibfnamefont {T.}~\bibnamefont {Gong}}, \bibinfo {author}
  {\bibfnamefont {R.}~\bibnamefont {Pan}}, \bibinfo {author} {\bibfnamefont
  {H.}~\bibnamefont {Wang}}, \bibinfo {author} {\bibfnamefont {Z.}~\bibnamefont
  {Guo}}, \bibinfo {author} {\bibfnamefont {H.}~\bibnamefont {Zhang}},\ and\
  \bibinfo {author} {\bibfnamefont {X.}~\bibnamefont {Fu}},\ }\bibfield
  {title} {\bibinfo {title} {{Recent advances in emerging Janus two-dimensional
  materials: from fundamental physics to device applications}},\ }\href@noop {}
  {\bibfield  {journal} {\bibinfo  {journal} {Journal of Materials Chemistry
  A}\ }\textbf {\bibinfo {volume} {8}},\ \bibinfo {pages} {8813} (\bibinfo
  {year} {2020}{\natexlab{b}})}\BibitemShut {NoStop}%
\bibitem [{\citenamefont {Yu}\ \emph {et~al.}(2021)\citenamefont {Yu},
  \citenamefont {Zhou}, \citenamefont {Zhang},\ and\ \citenamefont
  {Chang}}]{Yu2021}%
  \BibitemOpen
  \bibfield  {author} {\bibinfo {author} {\bibfnamefont {S.-B.}\ \bibnamefont
  {Yu}}, \bibinfo {author} {\bibfnamefont {M.}~\bibnamefont {Zhou}}, \bibinfo
  {author} {\bibfnamefont {D.}~\bibnamefont {Zhang}},\ and\ \bibinfo {author}
  {\bibfnamefont {K.}~\bibnamefont {Chang}},\ }\bibfield  {title} {\bibinfo
  {title} {{Spin Hall effect in the monolayer Janus compound MoSSe enhanced by
  Rashba spin-orbit coupling}},\ }\href
  {https://doi.org/10.1103/PhysRevB.104.075435} {\bibfield  {journal} {\bibinfo
   {journal} {Phys. Rev. B}\ }\textbf {\bibinfo {volume} {104}},\ \bibinfo
  {pages} {075435} (\bibinfo {year} {2021})}\BibitemShut {NoStop}%
\bibitem [{\citenamefont {Joseph}\ \emph {et~al.}(2021)\citenamefont {Joseph},
  \citenamefont {Roy},\ and\ \citenamefont {Narayan}}]{joseph2021}%
  \BibitemOpen
  \bibfield  {author} {\bibinfo {author} {\bibfnamefont {N.~B.}\ \bibnamefont
  {Joseph}}, \bibinfo {author} {\bibfnamefont {S.}~\bibnamefont {Roy}},\ and\
  \bibinfo {author} {\bibfnamefont {A.}~\bibnamefont {Narayan}},\ }\bibfield
  {title} {\bibinfo {title} {{Tunable topology and berry curvature dipole in
  transition metal dichalcogenide Janus monolayers}},\ }\href@noop {}
  {\bibfield  {journal} {\bibinfo  {journal} {Materials Research Express}\
  }\textbf {\bibinfo {volume} {8}},\ \bibinfo {pages} {124001} (\bibinfo {year}
  {2021})}\BibitemShut {NoStop}%
\bibitem [{\citenamefont {Guo}\ \emph {et~al.}(2021)\citenamefont {Guo},
  \citenamefont {Mu}, \citenamefont {Xiao},\ and\ \citenamefont
  {Liu}}]{guo2021}%
  \BibitemOpen
  \bibfield  {author} {\bibinfo {author} {\bibfnamefont {S.-D.}\ \bibnamefont
  {Guo}}, \bibinfo {author} {\bibfnamefont {W.-Q.}\ \bibnamefont {Mu}},
  \bibinfo {author} {\bibfnamefont {X.-B.}\ \bibnamefont {Xiao}},\ and\
  \bibinfo {author} {\bibfnamefont {B.-G.}\ \bibnamefont {Liu}},\ }\bibfield
  {title} {\bibinfo {title} {{Intrinsic room-temperature piezoelectric quantum
  anomalous hall insulator in Janus monolayer Fe 2 IX (X= Cl and Br)}},\
  }\href@noop {} {\bibfield  {journal} {\bibinfo  {journal} {Nanoscale}\
  }\textbf {\bibinfo {volume} {13}},\ \bibinfo {pages} {12956} (\bibinfo {year}
  {2021})}\BibitemShut {NoStop}%
\bibitem [{\citenamefont {Van~Thanh}\ \emph {et~al.}(2020)\citenamefont
  {Van~Thanh}, \citenamefont {Van}, \citenamefont {Saito}, \citenamefont {Hung}
  \emph {et~al.}}]{van2020}%
  \BibitemOpen
  \bibfield  {author} {\bibinfo {author} {\bibfnamefont {V.}~\bibnamefont
  {Van~Thanh}}, \bibinfo {author} {\bibfnamefont {N.~D.}\ \bibnamefont {Van}},
  \bibinfo {author} {\bibfnamefont {R.}~\bibnamefont {Saito}}, \bibinfo
  {author} {\bibfnamefont {N.~T.}\ \bibnamefont {Hung}}, \emph {et~al.},\
  }\bibfield  {title} {\bibinfo {title} {{First-principles study of mechanical,
  electronic and optical properties of Janus structure in transition metal
  dichalcogenides}},\ }\href@noop {} {\bibfield  {journal} {\bibinfo  {journal}
  {Applied Surface Science}\ }\textbf {\bibinfo {volume} {526}},\ \bibinfo
  {pages} {146730} (\bibinfo {year} {2020})}\BibitemShut {NoStop}%
\bibitem [{\citenamefont {Shi}\ and\ \citenamefont {Wang}(2018)}]{shi2018}%
  \BibitemOpen
  \bibfield  {author} {\bibinfo {author} {\bibfnamefont {W.}~\bibnamefont
  {Shi}}\ and\ \bibinfo {author} {\bibfnamefont {Z.}~\bibnamefont {Wang}},\
  }\bibfield  {title} {\bibinfo {title} {{Mechanical and electronic properties
  of Janus monolayer transition metal dichalcogenides}},\ }\href@noop {}
  {\bibfield  {journal} {\bibinfo  {journal} {Journal of Physics: Condensed
  Matter}\ }\textbf {\bibinfo {volume} {30}},\ \bibinfo {pages} {215301}
  (\bibinfo {year} {2018})}\BibitemShut {NoStop}%
\bibitem [{\citenamefont {Luo}\ \emph {et~al.}(2021)\citenamefont {Luo},
  \citenamefont {Han}, \citenamefont {Hu}, \citenamefont {Yuan}, \citenamefont
  {Jiao},\ and\ \citenamefont {Liu}}]{luo2021}%
  \BibitemOpen
  \bibfield  {author} {\bibinfo {author} {\bibfnamefont {Y.}~\bibnamefont
  {Luo}}, \bibinfo {author} {\bibfnamefont {S.}~\bibnamefont {Han}}, \bibinfo
  {author} {\bibfnamefont {R.}~\bibnamefont {Hu}}, \bibinfo {author}
  {\bibfnamefont {H.}~\bibnamefont {Yuan}}, \bibinfo {author} {\bibfnamefont
  {W.}~\bibnamefont {Jiao}},\ and\ \bibinfo {author} {\bibfnamefont
  {H.}~\bibnamefont {Liu}},\ }\bibfield  {title} {\bibinfo {title} {{The
  Thermal Stability of Janus Monolayers SnXY (X, Y= O, S, Se): Ab-Initio
  Molecular Dynamics and Beyond}},\ }\href@noop {} {\bibfield  {journal}
  {\bibinfo  {journal} {Nanomaterials}\ }\textbf {\bibinfo {volume} {12}},\
  \bibinfo {pages} {101} (\bibinfo {year} {2021})}\BibitemShut {NoStop}%
\bibitem [{\citenamefont {Tang}\ and\ \citenamefont {Kou}(2022)}]{tang2022}%
  \BibitemOpen
  \bibfield  {author} {\bibinfo {author} {\bibfnamefont {X.}~\bibnamefont
  {Tang}}\ and\ \bibinfo {author} {\bibfnamefont {L.}~\bibnamefont {Kou}},\
  }\bibfield  {title} {\bibinfo {title} {{2D Janus transition metal
  dichalcogenides: properties and applications}},\ }\href@noop {} {\bibfield
  {journal} {\bibinfo  {journal} {physica status solidi (b)}\ }\textbf
  {\bibinfo {volume} {259}},\ \bibinfo {pages} {2100562} (\bibinfo {year}
  {2022})}\BibitemShut {NoStop}%
\bibitem [{\citenamefont {Zhang}\ \emph {et~al.}(2022)\citenamefont {Zhang},
  \citenamefont {Xia}, \citenamefont {Li}, \citenamefont {Li}, \citenamefont
  {Fu}, \citenamefont {Cheng},\ and\ \citenamefont {Pan}}]{zhang2022}%
  \BibitemOpen
  \bibfield  {author} {\bibinfo {author} {\bibfnamefont {L.}~\bibnamefont
  {Zhang}}, \bibinfo {author} {\bibfnamefont {Y.}~\bibnamefont {Xia}}, \bibinfo
  {author} {\bibfnamefont {X.}~\bibnamefont {Li}}, \bibinfo {author}
  {\bibfnamefont {L.}~\bibnamefont {Li}}, \bibinfo {author} {\bibfnamefont
  {X.}~\bibnamefont {Fu}}, \bibinfo {author} {\bibfnamefont {J.}~\bibnamefont
  {Cheng}},\ and\ \bibinfo {author} {\bibfnamefont {R.}~\bibnamefont {Pan}},\
  }\bibfield  {title} {\bibinfo {title} {{Janus two-dimensional transition
  metal dichalcogenides}},\ }\href@noop {} {\bibfield  {journal} {\bibinfo
  {journal} {Journal of Applied Physics}\ }\textbf {\bibinfo {volume} {131}},\
  \bibinfo {pages} {230902} (\bibinfo {year} {2022})}\BibitemShut {NoStop}%
\bibitem [{\citenamefont {Rezavand}\ \emph {et~al.}(2022)\citenamefont
  {Rezavand}, \citenamefont {Ghobadi},\ and\ \citenamefont
  {Behnamghader}}]{rezavand2022electronic}%
  \BibitemOpen
  \bibfield  {author} {\bibinfo {author} {\bibfnamefont {A.}~\bibnamefont
  {Rezavand}}, \bibinfo {author} {\bibfnamefont {N.}~\bibnamefont {Ghobadi}},\
  and\ \bibinfo {author} {\bibfnamefont {B.}~\bibnamefont {Behnamghader}},\
  }\bibfield  {title} {\bibinfo {title} {Electronic and spintronic properties
  of janus m si 2 p x as y (m= mo, w) monolayers},\ }\href@noop {} {\bibfield
  {journal} {\bibinfo  {journal} {Physical Review B}\ }\textbf {\bibinfo
  {volume} {106}},\ \bibinfo {pages} {035417} (\bibinfo {year}
  {2022})}\BibitemShut {NoStop}%
\bibitem [{\citenamefont {Vojáček}\ \emph {et~al.}(2024)\citenamefont
  {Vojáček}, \citenamefont {Medina~Dueñas}, \citenamefont {Li},
  \citenamefont {Ibrahim}, \citenamefont {Manchon}, \citenamefont {Roche},
  \citenamefont {Chshiev},\ and\ \citenamefont {H}}]{vojacek2024field}%
  \BibitemOpen
  \bibfield  {author} {\bibinfo {author} {\bibfnamefont {L.}~\bibnamefont
  {Vojáček}}, \bibinfo {author} {\bibfnamefont {J.}~\bibnamefont
  {Medina~Dueñas}}, \bibinfo {author} {\bibfnamefont {J.}~\bibnamefont {Li}},
  \bibinfo {author} {\bibfnamefont {F.}~\bibnamefont {Ibrahim}}, \bibinfo
  {author} {\bibfnamefont {A.}~\bibnamefont {Manchon}}, \bibinfo {author}
  {\bibfnamefont {S.}~\bibnamefont {Roche}}, \bibinfo {author} {\bibfnamefont
  {M.}~\bibnamefont {Chshiev}},\ and\ \bibinfo {author} {\bibfnamefont
  {G.}~\bibnamefont {H}},\ }\bibfield  {title} {\bibinfo {title} {{Field-Free
  Spin-Orbit Torque Switching in Janus Chromium Dichalcogenides}},\ }\href@noop
  {} {\bibfield  {journal} {\bibinfo  {journal} {Nano Letters}\ } (\bibinfo
  {year} {2024})}\BibitemShut {NoStop}%
\bibitem [{\citenamefont {Liu}\ and\ \citenamefont {Shao}(2020)}]{liu2020two}%
  \BibitemOpen
  \bibfield  {author} {\bibinfo {author} {\bibfnamefont {Y.}~\bibnamefont
  {Liu}}\ and\ \bibinfo {author} {\bibfnamefont {Q.}~\bibnamefont {Shao}},\
  }\bibfield  {title} {\bibinfo {title} {Two-dimensional materials for
  energy-efficient spin--orbit torque devices},\ }\href@noop {} {\bibfield
  {journal} {\bibinfo  {journal} {ACS nano}\ }\textbf {\bibinfo {volume}
  {14}},\ \bibinfo {pages} {9389} (\bibinfo {year} {2020})}\BibitemShut
  {NoStop}%
\bibitem [{\citenamefont {Tang}\ \emph {et~al.}(2021)\citenamefont {Tang},
  \citenamefont {Liu}, \citenamefont {Li}, \citenamefont {Pan},\ and\
  \citenamefont {Zeng}}]{tang2021spin}%
  \BibitemOpen
  \bibfield  {author} {\bibinfo {author} {\bibfnamefont {W.}~\bibnamefont
  {Tang}}, \bibinfo {author} {\bibfnamefont {H.}~\bibnamefont {Liu}}, \bibinfo
  {author} {\bibfnamefont {Z.}~\bibnamefont {Li}}, \bibinfo {author}
  {\bibfnamefont {A.}~\bibnamefont {Pan}},\ and\ \bibinfo {author}
  {\bibfnamefont {Y.-J.}\ \bibnamefont {Zeng}},\ }\bibfield  {title} {\bibinfo
  {title} {Spin-orbit torque in van der waals-layered materials and
  heterostructures},\ }\href@noop {} {\bibfield  {journal} {\bibinfo  {journal}
  {Advanced Science}\ }\textbf {\bibinfo {volume} {8}},\ \bibinfo {pages}
  {2100847} (\bibinfo {year} {2021})}\BibitemShut {NoStop}%
\bibitem [{\citenamefont {Freimuth}\ \emph {et~al.}(2014)\citenamefont
  {Freimuth}, \citenamefont {Bl{\"u}gel},\ and\ \citenamefont
  {Mokrousov}}]{freimuth2014spin}%
  \BibitemOpen
  \bibfield  {author} {\bibinfo {author} {\bibfnamefont {F.}~\bibnamefont
  {Freimuth}}, \bibinfo {author} {\bibfnamefont {S.}~\bibnamefont
  {Bl{\"u}gel}},\ and\ \bibinfo {author} {\bibfnamefont {Y.}~\bibnamefont
  {Mokrousov}},\ }\bibfield  {title} {\bibinfo {title} {{Spin-orbit torques in
  Co/Pt (111) and Mn/W (001) magnetic bilayers from first principles}},\
  }\href@noop {} {\bibfield  {journal} {\bibinfo  {journal} {Physical Review
  B}\ }\textbf {\bibinfo {volume} {90}},\ \bibinfo {pages} {174423} (\bibinfo
  {year} {2014})}\BibitemShut {NoStop}%
\bibitem [{\citenamefont {Wang}\ \emph {et~al.}(2019)\citenamefont {Wang},
  \citenamefont {Meng}, \citenamefont {Zhang}, \citenamefont {Hou},
  \citenamefont {Finley}, \citenamefont {Han}, \citenamefont {Yang},\ and\
  \citenamefont {Liu}}]{wang2019}%
  \BibitemOpen
  \bibfield  {author} {\bibinfo {author} {\bibfnamefont {H.}~\bibnamefont
  {Wang}}, \bibinfo {author} {\bibfnamefont {K.-Y.}\ \bibnamefont {Meng}},
  \bibinfo {author} {\bibfnamefont {P.}~\bibnamefont {Zhang}}, \bibinfo
  {author} {\bibfnamefont {J.~T.}\ \bibnamefont {Hou}}, \bibinfo {author}
  {\bibfnamefont {J.}~\bibnamefont {Finley}}, \bibinfo {author} {\bibfnamefont
  {J.}~\bibnamefont {Han}}, \bibinfo {author} {\bibfnamefont {F.}~\bibnamefont
  {Yang}},\ and\ \bibinfo {author} {\bibfnamefont {L.}~\bibnamefont {Liu}},\
  }\bibfield  {title} {\bibinfo {title} {{Large spin-orbit torque observed in
  epitaxial SrIrO3 thin films}},\ }\href@noop {} {\bibfield  {journal}
  {\bibinfo  {journal} {Applied Physics Letters}\ }\textbf {\bibinfo {volume}
  {114}},\ \bibinfo {pages} {232406} (\bibinfo {year} {2019})}\BibitemShut
  {NoStop}%
\bibitem [{\citenamefont {Xie}\ \emph {et~al.}(2019)\citenamefont {Xie},
  \citenamefont {Lin}, \citenamefont {Yang}, \citenamefont {Shu}, \citenamefont
  {Chen}, \citenamefont {Liu}, \citenamefont {Yu}, \citenamefont {Breese},
  \citenamefont {Zhou}, \citenamefont {Yang} \emph {et~al.}}]{xie2019}%
  \BibitemOpen
  \bibfield  {author} {\bibinfo {author} {\bibfnamefont {Q.}~\bibnamefont
  {Xie}}, \bibinfo {author} {\bibfnamefont {W.}~\bibnamefont {Lin}}, \bibinfo
  {author} {\bibfnamefont {B.}~\bibnamefont {Yang}}, \bibinfo {author}
  {\bibfnamefont {X.}~\bibnamefont {Shu}}, \bibinfo {author} {\bibfnamefont
  {S.}~\bibnamefont {Chen}}, \bibinfo {author} {\bibfnamefont {L.}~\bibnamefont
  {Liu}}, \bibinfo {author} {\bibfnamefont {X.}~\bibnamefont {Yu}}, \bibinfo
  {author} {\bibfnamefont {M.~B.}\ \bibnamefont {Breese}}, \bibinfo {author}
  {\bibfnamefont {T.}~\bibnamefont {Zhou}}, \bibinfo {author} {\bibfnamefont
  {M.}~\bibnamefont {Yang}}, \emph {et~al.},\ }\bibfield  {title} {\bibinfo
  {title} {{Giant enhancements of perpendicular magnetic anisotropy and
  spin-orbit torque by a MoS2 layer}},\ }\href@noop {} {\bibfield  {journal}
  {\bibinfo  {journal} {Advanced Materials}\ }\textbf {\bibinfo {volume}
  {31}},\ \bibinfo {pages} {1900776} (\bibinfo {year} {2019})}\BibitemShut
  {NoStop}%
\bibitem [{\citenamefont {MacNeill}\ \emph {et~al.}(2017)\citenamefont
  {MacNeill}, \citenamefont {Stiehl}, \citenamefont {Guimaraes}, \citenamefont
  {Buhrman}, \citenamefont {Park},\ and\ \citenamefont {Ralph}}]{macneill2017}%
  \BibitemOpen
  \bibfield  {author} {\bibinfo {author} {\bibfnamefont {D.}~\bibnamefont
  {MacNeill}}, \bibinfo {author} {\bibfnamefont {G.}~\bibnamefont {Stiehl}},
  \bibinfo {author} {\bibfnamefont {M.}~\bibnamefont {Guimaraes}}, \bibinfo
  {author} {\bibfnamefont {R.}~\bibnamefont {Buhrman}}, \bibinfo {author}
  {\bibfnamefont {J.}~\bibnamefont {Park}},\ and\ \bibinfo {author}
  {\bibfnamefont {D.}~\bibnamefont {Ralph}},\ }\bibfield  {title} {\bibinfo
  {title} {{Control of spin--orbit torques through crystal symmetry in
  WTe2/ferromagnet bilayers}},\ }\href@noop {} {\bibfield  {journal} {\bibinfo
  {journal} {Nature Physics}\ }\textbf {\bibinfo {volume} {13}},\ \bibinfo
  {pages} {300} (\bibinfo {year} {2017})}\BibitemShut {NoStop}%
\bibitem [{\citenamefont {Husain}\ \emph
  {et~al.}(2020{\natexlab{b}})\citenamefont {Husain}, \citenamefont {Gupta},
  \citenamefont {Kumar}, \citenamefont {Kumar}, \citenamefont {Behera},
  \citenamefont {Brucas}, \citenamefont {Chaudhary},\ and\ \citenamefont
  {Svedlindh}}]{husain2020}%
  \BibitemOpen
  \bibfield  {author} {\bibinfo {author} {\bibfnamefont {S.}~\bibnamefont
  {Husain}}, \bibinfo {author} {\bibfnamefont {R.}~\bibnamefont {Gupta}},
  \bibinfo {author} {\bibfnamefont {A.}~\bibnamefont {Kumar}}, \bibinfo
  {author} {\bibfnamefont {P.}~\bibnamefont {Kumar}}, \bibinfo {author}
  {\bibfnamefont {N.}~\bibnamefont {Behera}}, \bibinfo {author} {\bibfnamefont
  {R.}~\bibnamefont {Brucas}}, \bibinfo {author} {\bibfnamefont
  {S.}~\bibnamefont {Chaudhary}},\ and\ \bibinfo {author} {\bibfnamefont
  {P.}~\bibnamefont {Svedlindh}},\ }\bibfield  {title} {\bibinfo {title}
  {{Emergence of spin--orbit torques in 2D transition metal dichalcogenides: A
  status update}},\ }\href@noop {} {\bibfield  {journal} {\bibinfo  {journal}
  {Applied Physics Reviews}\ }\textbf {\bibinfo {volume} {7}} (\bibinfo {year}
  {2020}{\natexlab{b}})}\BibitemShut {NoStop}%
\bibitem [{\citenamefont {Hanke}\ \emph {et~al.}(2017)\citenamefont {Hanke},
  \citenamefont {Freimuth}, \citenamefont {Niu}, \citenamefont {Bl{\"u}gel},\
  and\ \citenamefont {Mokrousov}}]{hanke2017mixed}%
  \BibitemOpen
  \bibfield  {author} {\bibinfo {author} {\bibfnamefont {J.-P.}\ \bibnamefont
  {Hanke}}, \bibinfo {author} {\bibfnamefont {F.}~\bibnamefont {Freimuth}},
  \bibinfo {author} {\bibfnamefont {C.}~\bibnamefont {Niu}}, \bibinfo {author}
  {\bibfnamefont {S.}~\bibnamefont {Bl{\"u}gel}},\ and\ \bibinfo {author}
  {\bibfnamefont {Y.}~\bibnamefont {Mokrousov}},\ }\bibfield  {title} {\bibinfo
  {title} {{Mixed Weyl semimetals and low-dissipation magnetization control in
  insulators by spin-orbit torques}},\ }\href@noop {} {\bibfield  {journal}
  {\bibinfo  {journal} {Nature Communications}\ }\textbf {\bibinfo {volume}
  {8}},\ \bibinfo {pages} {1479} (\bibinfo {year} {2017})}\BibitemShut
  {NoStop}%
\bibitem [{\citenamefont {Althammer}\ \emph {et~al.}(2015)\citenamefont
  {Althammer}, \citenamefont {Weiler}, \citenamefont {Huebl},\ and\
  \citenamefont {Goennenwein}}]{althammer2015spin}%
  \BibitemOpen
  \bibfield  {author} {\bibinfo {author} {\bibfnamefont {M.}~\bibnamefont
  {Althammer}}, \bibinfo {author} {\bibfnamefont {M.}~\bibnamefont {Weiler}},
  \bibinfo {author} {\bibfnamefont {H.}~\bibnamefont {Huebl}},\ and\ \bibinfo
  {author} {\bibfnamefont {S.~T.}\ \bibnamefont {Goennenwein}},\ }\bibfield
  {title} {\bibinfo {title} {Spin pumping},\ }\href@noop {} {\bibfield
  {journal} {\bibinfo  {journal} {Spintronics for Next Generation Innovative
  Devices}\ ,\ \bibinfo {pages} {111}} (\bibinfo {year} {2015})}\BibitemShut
  {NoStop}%
\bibitem [{\citenamefont {Shen}\ \emph {et~al.}(2014)\citenamefont {Shen},
  \citenamefont {Vignale},\ and\ \citenamefont
  {Raimondi}}]{shen2014microscopic}%
  \BibitemOpen
  \bibfield  {author} {\bibinfo {author} {\bibfnamefont {K.}~\bibnamefont
  {Shen}}, \bibinfo {author} {\bibfnamefont {G.}~\bibnamefont {Vignale}},\ and\
  \bibinfo {author} {\bibfnamefont {R.}~\bibnamefont {Raimondi}},\ }\bibfield
  {title} {\bibinfo {title} {{Microscopic theory of the inverse Edelstein
  effect}},\ }\href@noop {} {\bibfield  {journal} {\bibinfo  {journal}
  {Physical review letters}\ }\textbf {\bibinfo {volume} {112}},\ \bibinfo
  {pages} {096601} (\bibinfo {year} {2014})}\BibitemShut {NoStop}%
\bibitem [{\citenamefont {Freimuth}\ \emph {et~al.}(2017)\citenamefont
  {Freimuth}, \citenamefont {Bl\"ugel},\ and\ \citenamefont
  {Mokrousov}}]{freimuth2017}%
  \BibitemOpen
  \bibfield  {author} {\bibinfo {author} {\bibfnamefont {F.}~\bibnamefont
  {Freimuth}}, \bibinfo {author} {\bibfnamefont {S.}~\bibnamefont {Bl\"ugel}},\
  and\ \bibinfo {author} {\bibfnamefont {Y.}~\bibnamefont {Mokrousov}},\
  }\bibfield  {title} {\bibinfo {title} {{Charge pumping driven by the
  laser-induced dynamics of the exchange splitting}},\ }\href
  {https://doi.org/10.1103/PhysRevB.95.094434} {\bibfield  {journal} {\bibinfo
  {journal} {Phys. Rev. B}\ }\textbf {\bibinfo {volume} {95}},\ \bibinfo
  {pages} {094434} (\bibinfo {year} {2017})}\BibitemShut {NoStop}%
\bibitem [{\citenamefont {Wortmann}\ \emph {et~al.}(2023)\citenamefont
  {Wortmann}, \citenamefont {Michalicek}, \citenamefont {Baadji}, \citenamefont
  {Betzinger}, \citenamefont {Bihlmayer}, \citenamefont {Br\"oder},
  \citenamefont {Burnus}, \citenamefont {Enkovaara}, \citenamefont {Freimuth},
  \citenamefont {Friedrich}, \citenamefont {Gerhorst}, \citenamefont
  {Granberg~Cauchi}, \citenamefont {Grytsiuk}, \citenamefont {Hanke},
  \citenamefont {Hanke}, \citenamefont {Heide}, \citenamefont {Heinze},
  \citenamefont {Hilgers}, \citenamefont {Janssen}, \citenamefont
  {Kl\"uppelberg}, \citenamefont {Kovacik}, \citenamefont {Kurz}, \citenamefont
  {Lezaic}, \citenamefont {Madsen}, \citenamefont {Mokrousov}, \citenamefont
  {Neukirchen}, \citenamefont {Redies}, \citenamefont {Rost}, \citenamefont
  {Schlipf}, \citenamefont {Schindlmayr}, \citenamefont {Winkelmann},\ and\
  \citenamefont {Bl\"ugel}}]{fleur}%
  \BibitemOpen
  \bibfield  {author} {\bibinfo {author} {\bibfnamefont {D.}~\bibnamefont
  {Wortmann}}, \bibinfo {author} {\bibfnamefont {G.}~\bibnamefont
  {Michalicek}}, \bibinfo {author} {\bibfnamefont {N.}~\bibnamefont {Baadji}},
  \bibinfo {author} {\bibfnamefont {M.}~\bibnamefont {Betzinger}}, \bibinfo
  {author} {\bibfnamefont {G.}~\bibnamefont {Bihlmayer}}, \bibinfo {author}
  {\bibfnamefont {J.}~\bibnamefont {Br\"oder}}, \bibinfo {author}
  {\bibfnamefont {T.}~\bibnamefont {Burnus}}, \bibinfo {author} {\bibfnamefont
  {J.}~\bibnamefont {Enkovaara}}, \bibinfo {author} {\bibfnamefont
  {F.}~\bibnamefont {Freimuth}}, \bibinfo {author} {\bibfnamefont
  {C.}~\bibnamefont {Friedrich}}, \bibinfo {author} {\bibfnamefont {C.-R.}\
  \bibnamefont {Gerhorst}}, \bibinfo {author} {\bibfnamefont {S.}~\bibnamefont
  {Granberg~Cauchi}}, \bibinfo {author} {\bibfnamefont {U.}~\bibnamefont
  {Grytsiuk}}, \bibinfo {author} {\bibfnamefont {A.}~\bibnamefont {Hanke}},
  \bibinfo {author} {\bibfnamefont {J.-P.}\ \bibnamefont {Hanke}}, \bibinfo
  {author} {\bibfnamefont {M.}~\bibnamefont {Heide}}, \bibinfo {author}
  {\bibfnamefont {S.}~\bibnamefont {Heinze}}, \bibinfo {author} {\bibfnamefont
  {R.}~\bibnamefont {Hilgers}}, \bibinfo {author} {\bibfnamefont
  {H.}~\bibnamefont {Janssen}}, \bibinfo {author} {\bibfnamefont {D.~A.}\
  \bibnamefont {Kl\"uppelberg}}, \bibinfo {author} {\bibfnamefont
  {R.}~\bibnamefont {Kovacik}}, \bibinfo {author} {\bibfnamefont
  {P.}~\bibnamefont {Kurz}}, \bibinfo {author} {\bibfnamefont {M.}~\bibnamefont
  {Lezaic}}, \bibinfo {author} {\bibfnamefont {G.~K.~H.}\ \bibnamefont
  {Madsen}}, \bibinfo {author} {\bibfnamefont {Y.}~\bibnamefont {Mokrousov}},
  \bibinfo {author} {\bibfnamefont {A.}~\bibnamefont {Neukirchen}}, \bibinfo
  {author} {\bibfnamefont {M.}~\bibnamefont {Redies}}, \bibinfo {author}
  {\bibfnamefont {S.}~\bibnamefont {Rost}}, \bibinfo {author} {\bibfnamefont
  {M.}~\bibnamefont {Schlipf}}, \bibinfo {author} {\bibfnamefont
  {A.}~\bibnamefont {Schindlmayr}}, \bibinfo {author} {\bibfnamefont
  {M.}~\bibnamefont {Winkelmann}},\ and\ \bibinfo {author} {\bibfnamefont
  {S.}~\bibnamefont {Bl\"ugel}},\ }\href
  {https://doi.org/10.5281/zenodo.7576163} {\bibinfo {title} {{FLEUR}}},\
  \bibinfo {howpublished} {Zenodo} (\bibinfo {year} {2023})\BibitemShut
  {NoStop}%
\bibitem [{\citenamefont {Perdew}\ \emph {et~al.}(1996)\citenamefont {Perdew},
  \citenamefont {Burke},\ and\ \citenamefont
  {Ernzerhof}}]{perdew1996generalized}%
  \BibitemOpen
  \bibfield  {author} {\bibinfo {author} {\bibfnamefont {J.~P.}\ \bibnamefont
  {Perdew}}, \bibinfo {author} {\bibfnamefont {K.}~\bibnamefont {Burke}},\ and\
  \bibinfo {author} {\bibfnamefont {M.}~\bibnamefont {Ernzerhof}},\ }\bibfield
  {title} {\bibinfo {title} {{Generalized gradient approximation made
  simple}},\ }\href@noop {} {\bibfield  {journal} {\bibinfo  {journal}
  {Physical review letters}\ }\textbf {\bibinfo {volume} {77}},\ \bibinfo
  {pages} {3865} (\bibinfo {year} {1996})}\BibitemShut {NoStop}%
\bibitem [{\citenamefont {Shick}\ \emph {et~al.}(1999)\citenamefont {Shick},
  \citenamefont {Liechtenstein},\ and\ \citenamefont {Pickett}}]{Shick1999}%
  \BibitemOpen
  \bibfield  {author} {\bibinfo {author} {\bibfnamefont {A.~B.}\ \bibnamefont
  {Shick}}, \bibinfo {author} {\bibfnamefont {A.~I.}\ \bibnamefont
  {Liechtenstein}},\ and\ \bibinfo {author} {\bibfnamefont {W.~E.}\
  \bibnamefont {Pickett}},\ }\bibfield  {title} {\bibinfo {title}
  {{Implementation of the LDA+U method using the full-potential linearized
  augmented plane-wave basis}},\ }\href
  {https://doi.org/10.1103/PhysRevB.60.10763} {\bibfield  {journal} {\bibinfo
  {journal} {Phys. Rev. B}\ }\textbf {\bibinfo {volume} {60}},\ \bibinfo
  {pages} {10763} (\bibinfo {year} {1999})}\BibitemShut {NoStop}%
\bibitem [{\citenamefont {Kurz}(2002)}]{kurz2002}%
  \BibitemOpen
  \bibfield  {author} {\bibinfo {author} {\bibfnamefont {P.}~\bibnamefont
  {Kurz}},\ }\bibfield  {title} {\bibinfo {title} {{Magnetism and electronic
  structure of hcp Gd and the Gd (0001) surface}},\ }\href@noop {} {\bibfield
  {journal} {\bibinfo  {journal} {Journal of Physics: Condensed Matter}\
  }\textbf {\bibinfo {volume} {14}},\ \bibinfo {pages} {6353} (\bibinfo {year}
  {2002})}\BibitemShut {NoStop}%
\bibitem [{\citenamefont {Freimuth}\ \emph {et~al.}(2008)\citenamefont
  {Freimuth}, \citenamefont {Mokrousov}, \citenamefont {Wortmann},
  \citenamefont {Heinze},\ and\ \citenamefont {Bl\"ugel}}]{Freimuth2008}%
  \BibitemOpen
  \bibfield  {author} {\bibinfo {author} {\bibfnamefont {F.}~\bibnamefont
  {Freimuth}}, \bibinfo {author} {\bibfnamefont {Y.}~\bibnamefont {Mokrousov}},
  \bibinfo {author} {\bibfnamefont {D.}~\bibnamefont {Wortmann}}, \bibinfo
  {author} {\bibfnamefont {S.}~\bibnamefont {Heinze}},\ and\ \bibinfo {author}
  {\bibfnamefont {S.}~\bibnamefont {Bl\"ugel}},\ }\bibfield  {title} {\bibinfo
  {title} {{Maximally localized Wannier functions within the FLAPW
  formalism}},\ }\href {https://doi.org/10.1103/PhysRevB.78.035120} {\bibfield
  {journal} {\bibinfo  {journal} {Phys. Rev. B}\ }\textbf {\bibinfo {volume}
  {78}},\ \bibinfo {pages} {035120} (\bibinfo {year} {2008})}\BibitemShut
  {NoStop}%
\bibitem [{\citenamefont {Mostofi}\ \emph {et~al.}(2014)\citenamefont
  {Mostofi}, \citenamefont {Yates}, \citenamefont {Pizzi}, \citenamefont {Lee},
  \citenamefont {Souza}, \citenamefont {Vanderbilt},\ and\ \citenamefont
  {Marzari}}]{MOSTOFI20142309}%
  \BibitemOpen
  \bibfield  {author} {\bibinfo {author} {\bibfnamefont {A.~A.}\ \bibnamefont
  {Mostofi}}, \bibinfo {author} {\bibfnamefont {J.~R.}\ \bibnamefont {Yates}},
  \bibinfo {author} {\bibfnamefont {G.}~\bibnamefont {Pizzi}}, \bibinfo
  {author} {\bibfnamefont {Y.-S.}\ \bibnamefont {Lee}}, \bibinfo {author}
  {\bibfnamefont {I.}~\bibnamefont {Souza}}, \bibinfo {author} {\bibfnamefont
  {D.}~\bibnamefont {Vanderbilt}},\ and\ \bibinfo {author} {\bibfnamefont
  {N.}~\bibnamefont {Marzari}},\ }\bibfield  {title} {\bibinfo {title} {{An
  updated version of wannier90: A tool for obtaining maximally-localised
  Wannier functions}},\ }\href
  {https://doi.org/https://doi.org/10.1016/j.cpc.2014.05.003} {\bibfield
  {journal} {\bibinfo  {journal} {Computer Physics Communications}\ }\textbf
  {\bibinfo {volume} {185}},\ \bibinfo {pages} {2309} (\bibinfo {year}
  {2014})}\BibitemShut {NoStop}%
\end{thebibliography}%

\end{document}